\documentclass[journal]{IEEEtran}
\usepackage{cite}

\ifCLASSINFOpdf
\else
\fi

\usepackage[cmex10]{amsmath}

\usepackage{amsfonts}
\usepackage[ruled,vlined]{algorithm2e}
\usepackage{algcompatible}
\usepackage{subfigure}
\usepackage{graphicx}
\usepackage{xcolor}
\usepackage{multirow}
\usepackage{amssymb}
\usepackage{array}
\usepackage{txfonts}
\usepackage{bm}
\usepackage{arydshln}
\usepackage{makecell}

\newcommand{\PreserveBackslash}[1]{\let\temp=\\#1\let\\=\temp}

\newcolumntype{C}[1]{>{\PreserveBackslash\centering}p{#1}}
\newcolumntype{R}[1]{>{\PreserveBackslash\raggedleft}p{#1}}
\newcolumntype{L}[1]{>{\PreserveBackslash\raggedright}p{#1}}

\usepackage[pagebackref=false,breaklinks=true,letterpaper=true,colorlinks=false,bookmarks=false]{hyperref}

\begin{document}

%

\title{SGUIE-Net: Semantic Attention Guided Underwater Image Enhancement with Multi-Scale Perception}
%
%
%

\author{Qi Qi,~\IEEEmembership{Student Member,~IEEE},
        Kunqian Li*,~\IEEEmembership{Member,~IEEE},
        Haiyong Zheng,~\IEEEmembership{Member,~IEEE},
        Xiang Gao,
        Guojia Hou, ~\IEEEmembership{Member,~IEEE},
        Kun Sun, ~\IEEEmembership{Member,~IEEE}
\thanks{Q. Qi and H. Zheng are with Faculty of Information Science and Engineering, Ocean University of China, Qingdao 266100, China (qiqi2013@stu.ouc.edu.cn; zhenghaiyong@ouc.edu.cn).}
\thanks{K. Li and X. Gao are with College of Engineering, Ocean University of China, Qingdao 266100, China (likunqian@ouc.edu.cn; xgao@ouc.edu.cn).}
\thanks{G. Hou is with the College of Computer Science and Technology, Qingdao University, Qingdao 266071, China (hgjouc@126.com).}
\thanks{K. Sun is with School of Computer Science, China University of Geosciences, Wuhan 430078, China (sunkun@cug.edu.cn).}
\thanks{The research has been supported by the National Natural Science Foundation of China under Grant 61906177, in part by the National Natural Science Foundation of China under Grant 62176242, in part by the Natural Science Foundation of Shandong Province under Grant ZR2019BF034, in part by Significant Applied Technology Innovation Projects for Agriculture of Shandong Province under Grants SD2019NJ020, and in part by Fundamental Research Funds for the Central Universities under Grants 201964013.}
\thanks{$^{*}$ Corresponding author: Kunqian Li (likunqian@ouc.edu.cn)}

}

\maketitle

\begin{abstract}
  Due to the wavelength-dependent light attenuation, refraction and scattering, underwater images usually suffer from color distortion and blurred details. However, due to the limited number of paired underwater images with undistorted images as reference, training deep enhancement models for diverse degradation types is quite difficult. 
To boost the performance of data-driven approaches, it is essential to establish more effective learning mechanisms that mine richer supervised information from limited training sample resources.
  In this paper, we propose a novel underwater image enhancement network, called SGUIE-Net, in which we introduce semantic information as high-level guidance across different images that share common semantic regions. Accordingly, we propose semantic region-wise enhancement module to perceive the degradation of different semantic regions from multiple scales and feed it back to the global attention features extracted from its original scale. 
This strategy helps to achieve robust and visually pleasant enhancements to different semantic objects, which should thanks to the guidance of semantic information for differentiated enhancement. More importantly, for those degradation types that are not common in the training sample distribution, the  guidance connects them with the already well-learned types according to their semantic relevance. 
  Extensive experiments on the publicly available datasets and our proposed dataset demonstrated the impressive performance of SGUIE-Net. The code and proposed dataset are available at: \url{https://trentqq.github.io/SGUIE-Net.html}
  \end{abstract}
  
  
  %
  \IEEEpeerreviewmaketitle
  
  \section{Introduction}\label{sec:Introduction}
  \IEEEPARstart{U}Nmanned underwater robots equipped with high-performance vision modules have been increasingly becoming a major tool for ocean exploration. However, the underwater imaging environment is quite complicated, and is dynamically changed by local disturbances, underwater floating sand, plankton, illumination differences, etc. In addition, underwater images are often characterized by low contrast and blurred detail information due to the effect of underwater light attenuation, which brings great challenges to vision-based underwater operations \cite{Gao2019Underwater}. Therefore, in recent years, underwater image enhancement have received extensive attention and intensive research to improve the visual quality of underwater images \cite{Yang2019InDepthSurvey}. 
  
  With the great success of deep learning in many areas of computer vision, it also brings new strategies and perspectives to underwater image enhancement task \cite{ANWAR2020Diving}. However, it is almost impossible to simultaneously obtain both degraded and clear underwater images in situ, which makes it difficult to obtain a large number of paired samples to train a completed deep enhancement model suitable for diverse degradation types. As a compromise, researchers proposed to use synthetic underwater datasets \cite{Fabbri2018UGAN, LI2020Underwater} or manually selected enhancements with optimal subjective visual quality as model training references \cite{Li2020UIEBD, Qi2021Underwater}. Unfortunately, for the former strategy, there is still a clear gap between synthetic underwater images and real underwater images in terms of verisimilitude and scene diversity. As to the manually selected references, the labelling is costly in labor and they are not veritable ground-truth, which still carry clear human subjective visual preference and may introduce label contradiction for similar underwater scenarios. 
  
  \begin{figure}[t]
    \begin{center}
    \includegraphics[width=1\linewidth]{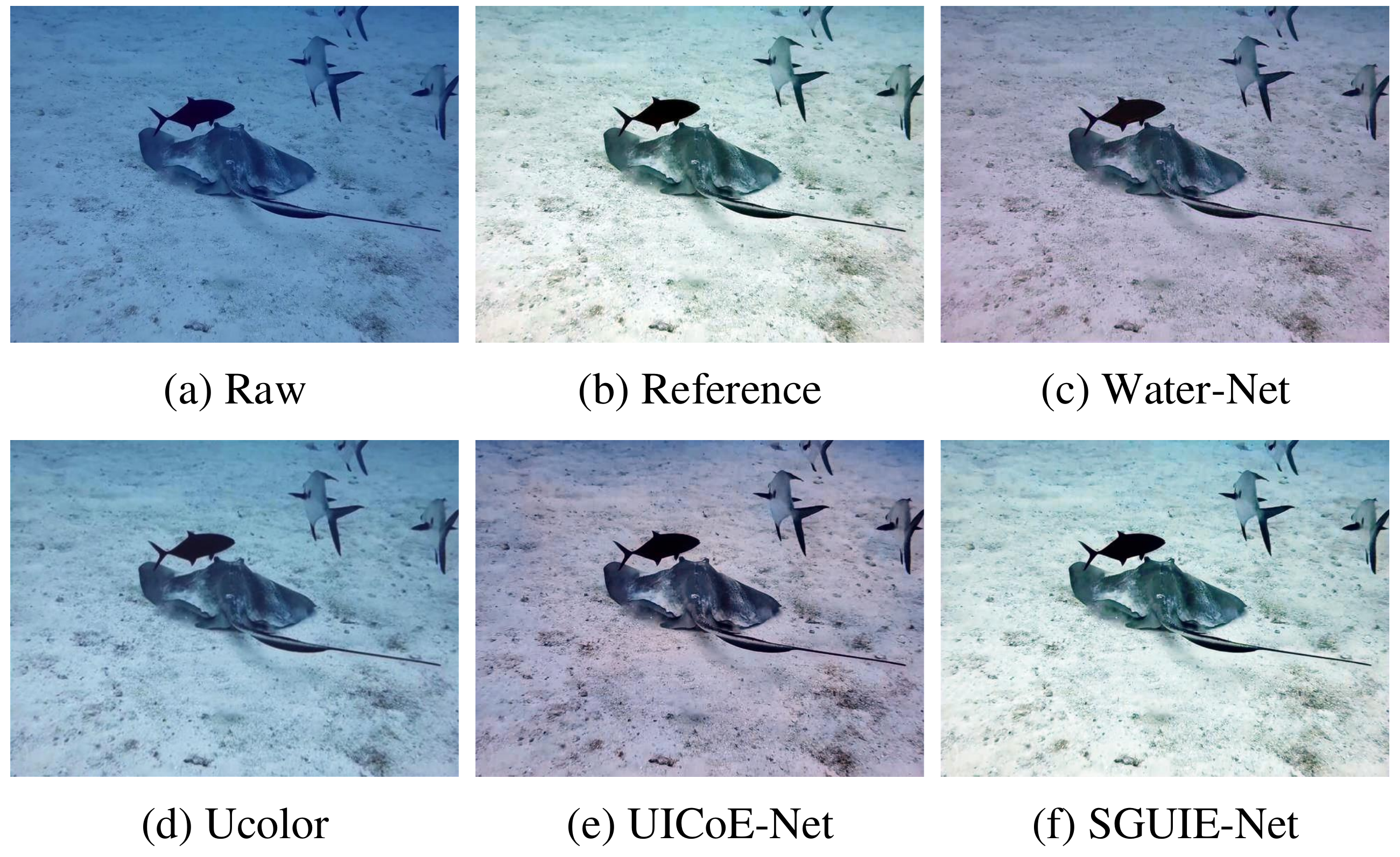}
    \caption{Enhancement example of deep models with and without semantic guidance. (a) and (b) are the raw underwater image and reference enhancement. (c)-(e) present enhancements of Water-Net \cite{Li2020UIEBD}, Ucolor \cite{li2021underwater} and UICoE-Net \cite{Qi2021Underwater}, which has no semantic guidance. (f) is the enhancement of the proposed SGUIE-Net which introduces semantic guidance.}
    \label{fig:intro}
    \end{center}
  \end{figure}
  
  Based on the above analysis, we realize that establishing more effective learning mechanisms is quite important for the mining of reliable supervisory information and robust enhancement. Collecting supervisory information from more reliable labelled data as auxiliary constraint, such as semantic maps, is a promising strategy. The semantic map not only portrays the semantic region boundaries, but also establishes high-level semantic associations between images, which is important for improving the consistency of enhanced appearance of semantic related  regions and improving the model generalization during the training process. In addition, as introduced in \cite{XIE2020Semantically}, explicitly harnessing the scene semantics into the enhancement process can suppress the introduction of artifacts, such as boosted noises and unnatural visual appearances.
  
  In this paper, in order to leverage semantic maps to assist in underwater image enhancement task, we propose an underwater image enhancement network with semantic attention guidance and multi-scale perception, whose corresponding deep model is named as Semantic attention Guided Underwater Image Enhancement Network (SGUIE-Net). Figure \ref{fig:intro} shows a comparison of enhancements with and without semantic guidance. For those correction mappings that are not effectively learned and portrayed by traditional deep models, the semantic guidance introduced into SGUIE-Net can provide effective auxiliary constraints to fix such unsatisfactory enhancement. The contributions of this paper are summarized as follows
  \begin{itemize}
    \item [1)]
    We introduce semantic association into underwater image enhancement task and propose a semantic attention guided underwater image enhancement network called SGUIE-Net. It learns features incorporating high-level semantic information to build enhancement guidance for those degradation types that are uncommon in the training sample distribution but semantically relevant with the well-learned types.
    \item [2)]
    We design SGUIE-Net as a deep enhancement network with multi-scale perception, whose main branch and semantic branch are fused to complement each other. The main branch is used to provide end-to-end enhancement while preserving image texture details in original scale, and the semantic branch is used to complement the semantically guided features with multi-scale perception. 
    \item [3)]
    We establish a new benchmark, namely SUIM-E, by extending the Segmentation of Underwater IMagery (SUIM) dataset \cite{islam2020semantic} with corresponding enhancement reference images. Then, comprehensive experiments, evaluations and analyses conducted on multiple datasets verify the good performance of the proposed SGUIE-Net.
  \end{itemize}

  \section{Related Works}\label{sec:relatedworks}
  
  \subsection{Deep Enhancement Models Trained with Paired Samples}
  As deep learning has made great strides in many low-level vision tasks, more and more researchers have started to introduce deep learning into the field of underwater image enhancement \cite{ANWAR2020Diving}. Learning strategies with real-world paired samples have also received widespread attention. In the last few years, researchers have made extensive effort on new sample generation strategies, more effective learning strategies and network architectures. 
  
  Thanks to the great success of image transfer techniques, Fabbri et al. \cite{Fabbri2018UGAN} proposed to synthesize paired underwater images with CycleGAN \cite{Zhu2017CycleGAN}. Deep models that use such samples also includes FGAN \cite{li2019fusion}, DenseGAN \cite{Guo2020DenseGAN}, MLFcGAN \cite{liu2019mlfcgan} and FUnIE-GAN \cite{Islam2020EUVPFUnIEGAN}. Later, Li et al. \cite{Li2020UIEBD} constructed a real-world underwater image enhancement benchmark (UIEB), which contains 890 paired real-world underwater images. The paired references are manually selected from enhancement candidates according to subject visual preference. Although the above strategies have expanded the sources of paired data, there is still a shortage of high-quality training samples that truly match real-world underwater scenarios and diverse degradation.
  
  To make full use of the limited high-quality underwater training images, researchers also explored new learning strategies and network structures. With UIEB dataset, Li et al. \cite{Li2020UIEBD} proposed a gated fusion network called Water-Net, which fuses the inputs with three predicted confidence maps to achieve the enhanced result. Qi et al. \cite{Qi2021Underwater} proposed an Underwater Image Co-enhancement Network (UICoE-Net) by introducing correlation feature matching units to communicate the mutual correlation of the two input branches. More recently, Li et al. \cite{li2021underwater} proposed to learn rich feature representations from diverse color spaces and attention weights by medium transmission map, and accordingly designed an encode-decoder enhancement network called Ucolor.
  
  While, the shortage of high quality paired training samples has always constrained the performance of the above models. Although some researchers have started to explore shared features among images for richer constraints, such as UICoE-Net, there is still a gap in the mining of high-level constraints oriented to scene-consistent enhancement. And such non-consistency within the same scene is often caused by the inherent ambiguity of the training labels. To address this concern, we propose to introduce semantic guidance as high-level constraint for semantically robust enhancement.

  \subsection{Semantic Guidance in Computer Vision}
  In previous studies, semantic guidance has been widely explored in various computer vision tasks, such as visual navigation \cite{Visual2019Mousavian, Wu2019Bayesian}, image generation \cite{Hong2018Inferring, Tang2020Local}, image-image matching \cite{Choy2016Universal, GLU-Net2020Truong}, image-sentence matching \cite{Image2020Huang, Xu2020Cross}, Re-IDentification \cite{Kalayeh2018Human, Zhang2019Densely}, etc. Most of these tasks are high-level or mid-level tasks, where semantic information plays a guiding and error-correcting role. A common form is that the result of the associated task is constrained by semantic information, and the result violating the constraint will be penalized by extra cost \cite{Choy2016Universal, GLU-Net2020Truong, Tang2020Local, Image2020Huang, Xu2020Cross}. Another form is that semantic information will be used as prior knowledge to provide guidance for subsequent tasks, thus providing additional constraints while reducing the search space \cite{Visual2019Mousavian, Wu2019Bayesian, Hong2018Inferring, Kalayeh2018Human, Zhang2019Densely}. 
  
  \begin{figure*}[t]
    \begin{center}
    \includegraphics[width=1\linewidth]{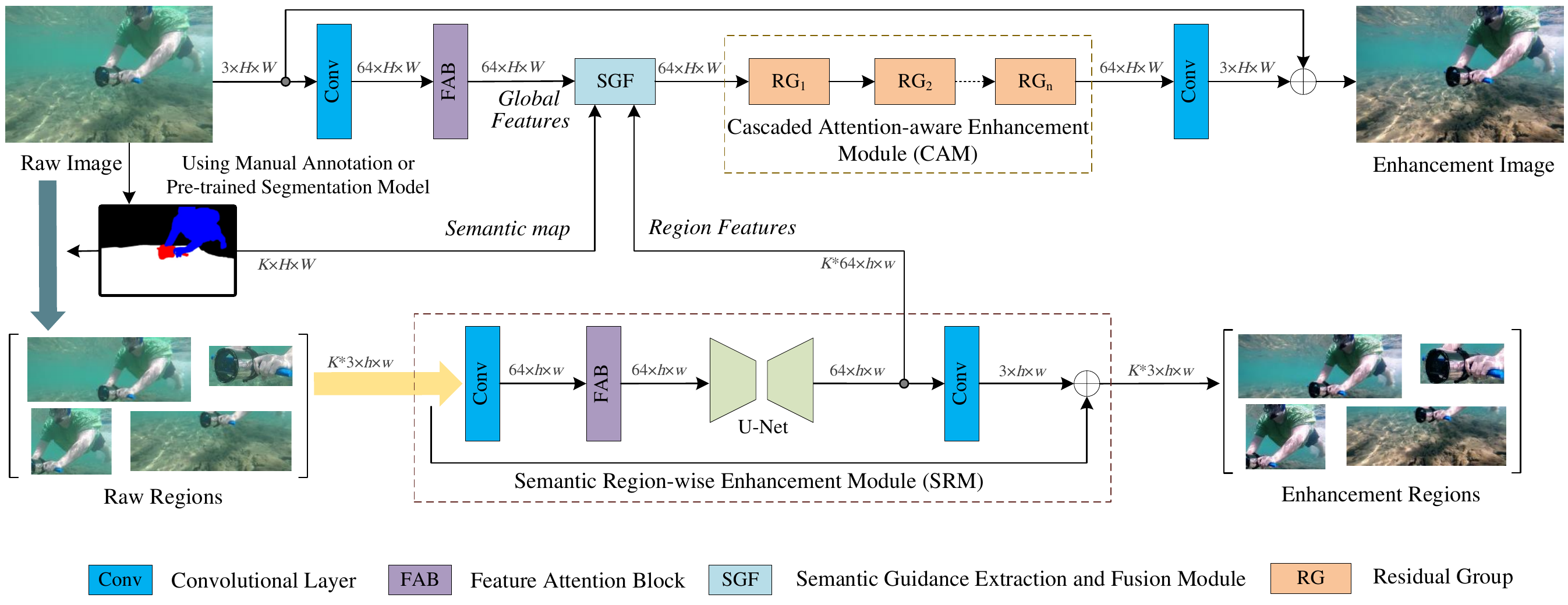}
    \caption{Architecture of SGUIE-Net. Our SGUIE-Net contains two enhancement branches for multi-scale perception, the main (top) branch works on the original input scale through the cascaded attention-aware enhancement module which contains multiple residual groups (RGs), and additionally embeds a semantic guided feature extraction and fusion module (SGF) to achieve enhanced multi-scale features. The bottom branch works on the semantic level, learning multi-scale semantic association features with encoder-decoder structure. It builds semantic enhancement guidance through the semantic region-wise enhancement module (SRM) and then feeds them back to the main branch. The details of the FAB, RG and SGF modules are shown in Figure \ref{fig:FFB}.}
    \label{fig:SGUIE}
    \end{center}
  \end{figure*}
  
  In addition to the above tasks, we also notice that semantic information is starting to play an important role in some of the low-level image processing tasks, such as image enhancement \cite{Lindner2012Joint, XIE2020Semantically}, restoration \cite{Liu2019Coherent} and manipulation \cite{Pan2020Exploiting}. As an early attempt in semantic image enhancement area, Lindner and Shaji \cite{Lindner2012Joint} established an non-parametric statistical framework to link the image characteristics and semantic concepts. Then the semantic context is used to guide the gray-level tone-mapping, emphasizing semantically relevant
  colors, and performing a defocus magnification for the given image. Xie et al. \cite{XIE2020Semantically} incorporated semantic information into a fusion-based low-light image enhancement framework. Liu et al. \cite{Liu2019Coherent} proposed a deep generative model for image inpainting with a coherent semantic attention layer, which can help to preserve contextual structure and model the semantic relevance between the content-missing holes. Pan et al. \cite{Pan2020Exploiting} exploited rich image prior, including high-level concepts, captured by a generative adversarial network for diverse image manipulation. For underwater image enhancement, the semantic information provides a bridge to link different samples for robust model training, which facilitates the sharing of supervisory information for limited samples and also improves the consistency of the enhancement of regions with same semantic attribute.
  
  \section{Proposed Method}
  \subsection{Overall Network with Multi-Scale Perception}
  
  The architecture of the proposed SGUIE-Net is shown in Figure \ref{fig:SGUIE}. With semantic information as a high-level and cross-image constraint, it provides a new perspective to more effectively mine limited supervisory information.
  By manually performing mask annotation or using pre-trained underwater semantic segmentation models, such as SUIM-Net \cite{islam2020semantic}, we partition the input image into sub-regions containing the corresponding semantic objects. Then, we deliver them to the semantic region-wise enhancement module, which uses an encoder-decoder structure to extract multi-scale contextual information and semantic attention features. 
  Subsequently, the semantic guidance extraction and fusion module fuses these semantic-aware local features to the global features in original resolution according to the guidance of the semantic masks. 
  Since the encoder-decoder structure prone to lose spatial detail features, the main branch (top branch in Figure \ref{fig:SGUIE}) uses a cascaded attention-aware enhancement module, running at the original scale to provide differentiated enhancement for uneven degradation and maintain the detail texture.
  
  \subsection{Cascaded Attention-aware Enhancement Module}\label{sec:CAM}
  
  Although the existing single-branch enhancement methods based on encoder-decoder architecture can well capture multi-scale features with multiple downsampling operations, they also inevitably lose pixel-wise spatial details and dependency in high-level features. To well preserve the detailed texture appearance of the input image, we design a cascaded attention-aware enhancement module (CAM) shown in Figure \ref{fig:SGUIE} as the basic structure of the main branch. Note that the cascaded attention-aware module does not contain any down-sampling or up-sampling operations, and the convolutional layers used in this module are all set with ${kernel\_size=3}$ and ${padding=1}$ to keep the feature scale fixed during the process in the main branch and to maximally preserve spatial detail features.
  
\begin{figure*}[htb]
    \begin{center}
    \includegraphics[width=1\linewidth]{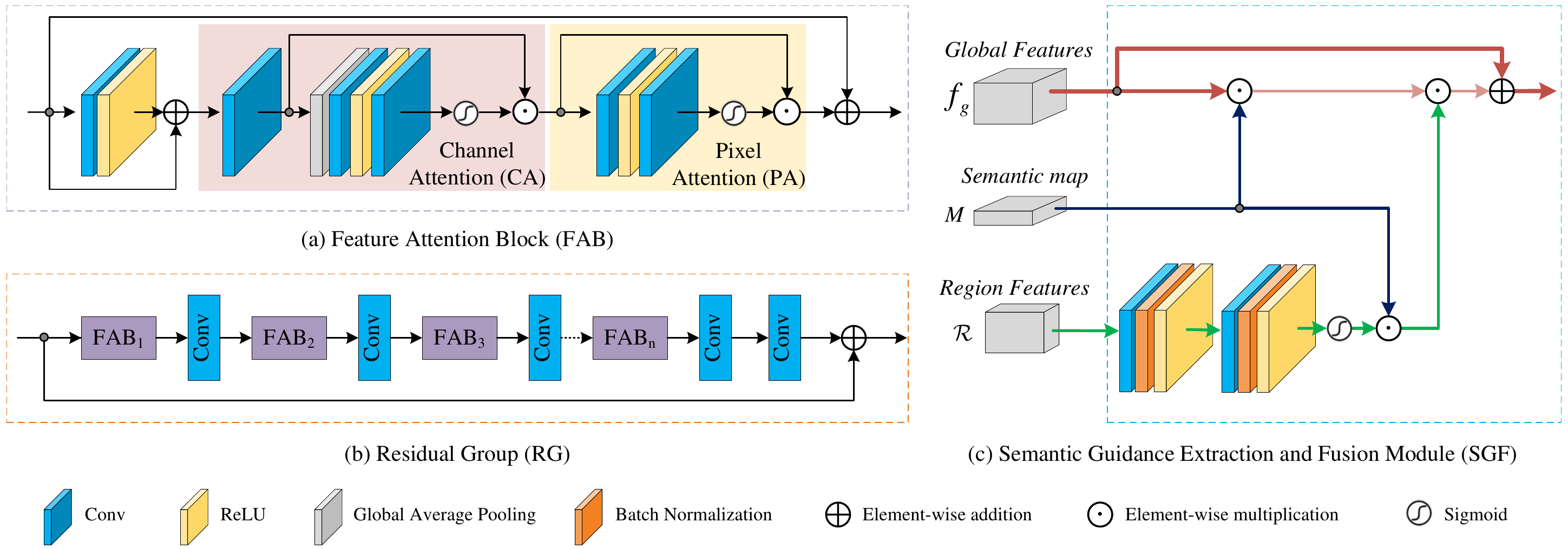}
    \caption{The architecture of the basic blocks of our SGUIE-Net. (a) Feature attention block (FAB) consists of cascaded channel attention (CA) and pixel attention (PA) module. (b) illustrates the RG module formed by multiple FABs with shorted connections. (c) The SFG module takes the global attention features obtained by the head of the main branch, semantic mask and the semantic region-based attention features as the inputs. Then, the SFG module extracts the local parts of the global attention features according to the semantic guidance for the later fusion on the semantic region-based attention features.}
    \label{fig:FFB}
    \end{center}
  \end{figure*}

CAM consists of three sets of cascaded residual groups (RGs) with skip connections as shown in Figure \ref{fig:FFB}(b). In the RG module, we stack four feature attention blocks (FAB) as shown in Figure \ref{fig:FFB}(a) to learn higher-level features, and use long skip connections to solve the problem of training difficulties caused by the network being too deep. Considering that our cascaded attention-aware enhancement module is designed to perceive the image features at the original image scale, we expect to further obtain differentiated features that fully describe the uneven degradation of the input image. Inspired by \cite{qin2020ffa}, FAB consists of a channel attention module (CA) and a pixel attention module (PA) placed in sequence. The detailed discussion of whether to use series or parallel connections for the two modules can be found in \cite{woo2018cbam}.
  
  \subsection{Semantic Region-wise Enhancement Module}\label{sec:SRM}
  
  One of the most challenging problems currently facing the underwater enhancement task is the lack of realistic reference corresponding to degraded images, which prevents various existing deep learning networks from achieving better performance through effective data driving. 
  We have to find other effective and economical alternatives to provide supplementary information. Semantic segmentation as an important fundamental task of computer vision is to understand images from the pixel level. Semantic maps contain a rich collection of high-level semantic information. We would like to make the enhancement network learn association features with high-level semantic information to build supplementary enhancement guidance for unusual cases from diverse degradation. 
  
  Benefiting from the fact that the CAM maintains the original resolution of the input image, we can more comfortably propose a semantic region-wise enhancement module (SRM) for regional multi-scale perception. With the subsequently mentioned SGF module, these multi-scale features containing perceptual feedback for different semantic regions can be fused to the main-branch features with original resolution. It enables the CAM to perform differentiated enhancement for semantic regions with different perceptual features. Moreover, SRM establishes high-level guidance for those degradation types that are uncommon in the training sample distribution but semantically relevant with the well-learned types.

  The input of the SRM is a collection of sub-images containing different semantic regions, which are obtained by splitting the original input image according to the semantic mask. The mask can be either manually annotated or generated with a pre-trained underwater semantic segmentation model, such as SUIM-Net \cite{islam2020semantic}. For convenience, we denote the raw input image as $I$, then $I_{k}\in I,k=1,2,\ldots,K$ represents the $k$-th semantic region of $I$ and $K$ is the number of semantic categories contained in $I$. Each input region $I_{k}$ is first passed through the convolutional layer and the FAB module to learn the residuals, and then a preliminary pre-processed feature denoted as $\mathcal{S}_k$ is obtained by a shortcut connection as follows
  \begin{equation}
    {\mathcal{S}_k = I_k \oplus C_{FAB} (C_{conv}(I_k))},
  \end{equation}
  where $C_{FAB}$ represents the processing of the FAB module and $ C_{conv} $ denotes the convolution operation. 
  Since the U-Net-like network structure has been widely used in underwater enhancement missions due to its ability to obtain a larger range of multi-scale information \cite{Qi2021Underwater, li2021underwater}, we also use the standard U-Net structure \cite{ronneberger2015u} for the enhancement and multi-scale feature transform of these raw semantic regions. Unlike before, what we learn through the U-Net module is the residuals of the enhanced regions denoted as 
  \begin{equation}
    {\mathcal{R}_k = C_{U}(\mathcal{S}_k)},
  \end{equation}
  where $C_{U}$ represents the processing of U-Net module.
  Then, the enhancement regions will be acquired as follows
  \begin{equation}
    {E_k = I_k \oplus C_{conv}(\mathcal{R}_k)},
  \end{equation}
  where $E_k$ represents the enhancement for region $I_k$.
  Note that the $\mathcal{R}$ learned by U-Net module contain multi-scale attention features for regions with different semantic attributes. Therefore, we next feed these features into the main branch to inject these semantic guidance for high-level and cross-image enhancement alignment.
  
  \subsection{Semantic Guidance Extraction and Fusion Module}\label{sec:SGF}
   
  In the proposed network, the CAM misses multi-scale information since it is designed to preserve pixel-wise spatial details and dependency at the original resolution, while the SRM acquires a amount of significant multi-scale semantic attention information for different semantic regions. In order to fuse these two kinds of features, we design the semantic guidance extraction and fusion module (SGF). As shown in Figure \ref{fig:FFB}(c), the SGF module has three inputs, one is the global attention feature of the raw image denoted as $f_g \in \mathbb{R}^{C \times H \times W}$, another is the residual features of semantic regions denoted as $\mathcal{R} \in \mathbb{R}^{C \times h \times w}$ where $h \in [0, H]$ and $w \in [0,W]$, and the last one is the semantic region mask denoted as $m \in \mathbb{R}^{1 \times h \times w}$ which is the part of $M \in \mathbb{R}^{K \times H \times W}$.

\begin{figure}[t]
    \begin{center}
    \includegraphics[width=1\linewidth]{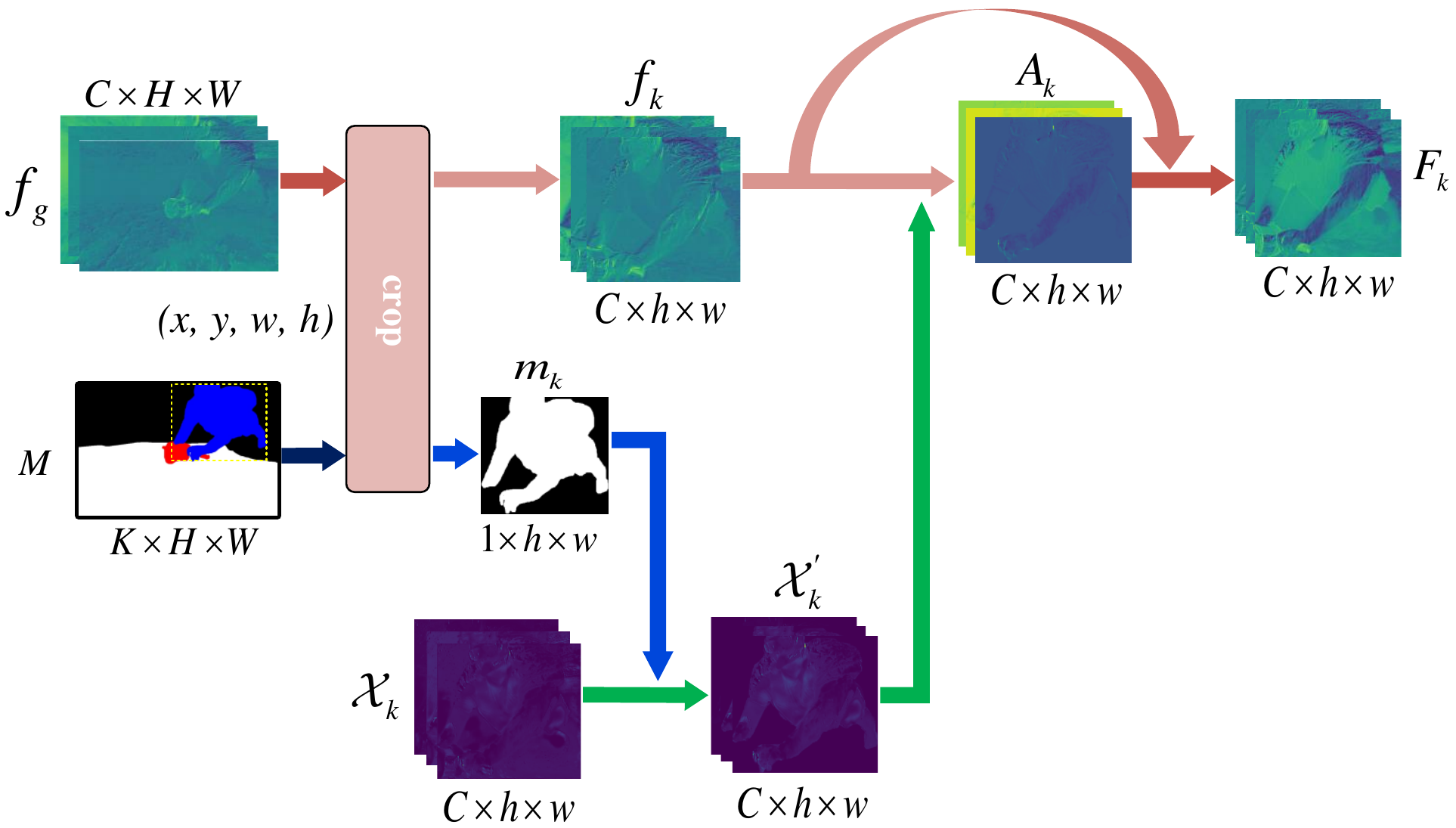}
    \caption{Diagram of fusing the semantic region attention features.}
    \label{fig:SGF}
    \end{center}
  \end{figure}

  A visual example of the feature fusion is given in Figure \ref{fig:SGF}. For the residual features $\mathcal{R}$ extracted by the U-Net network module, it first passes through a stack of double Conv-BN-ReLU units and a sigmoid function. Then, the multi-scale semantic region attention feature $\mathcal{X}_k$ can be acquired as follows
  \begin{equation}
    {\mathcal{X}_k = \sigma (C_{Dconv}(\mathcal{R}_k))},
  \end{equation}
  where $C_{Dconv}$ denotes the double Conv-BN-ReLU units, $k \in \{1,2,\ldots,K\}$ represents the $k$-th region and $ \sigma $ represents the sigmoid function. As shown in Figure \ref{fig:SGUIE}, the semantic regions fed into SRM are the largest outer rectangle of the actual semantic object in order to increase the context-awareness. While, in the fusion stage, we need to eliminate the redundant surrounding information and only keep the attention-aware features of the semantic regions. 
  As shown in Figure \ref{fig:SGF}, given a semantic region mask $m_k$ and the semantic attention feature $\mathcal{X}_k$, the regional semantic attention features $\mathcal{X}_{k}^{\prime}$ is defined as
  \begin{equation}
    {\mathcal{X}_{k}^{\prime} = \mathcal{X}_k \odot m_k}.
  \end{equation}
  In order to make the semantic regional attention features can be smoothly incorporated into the global features, we extract the corresponding part of the global features denoted as $f_k$ in advance by using the location information of the semantic regions. Then the cropped feature from main branch, which has been highlighted by semantic attention can be acquired as
  \begin{equation}
    {A_k = f_k \odot \mathcal{X}_{k}^{\prime}}.
  \end{equation}
With element-wise addition for further fusion, the $k$-th regional feature of the main branch can be denoted as
  \begin{equation}
    {F_k = f_k \oplus A_k}.
  \end{equation}
Finally, the global features that injected with semantic guidance will be sent to the CAM for residual perception and final enhancement. 
    
  \subsection{Loss Function}\label{sec:PCC}
  
  Inspired by \cite{LI2020Underwater,Qi2021Underwater}, to better preserve the sharpness of edges and details in enhancement images, we use the $\ell_2$ loss to minimize the pixel-wise error.
  The $\ell_2$ loss is defined as follows
  \begin{equation}
    {\mathcal{L} = \sum_{i=1}^{H}\sum_{j=1}^{W} \left( E(i,j) - {E_{GT}}(i,j) \right)^2},
  \end{equation}
  where the $E$ and $E_{GT}$ represent the enhancement of input image $I$  and the ground truth image, respectively. $i$ and $j$ are the pixel location indexes. 
  
  \section{Experiments}

  In this section, we first introduce the experimental settings and the SUIM-E datasets\footnote{https://drive.google.com/drive/folders/1gA3Ic7yOSbHd3w214-AgMI9UleAt4bRM?usp=sharing} we established, then we compare our method with state-of-the-arts on six datasets. Finally, series of ablation studies are conducted to demonstrate the effectiveness of each component in SGUIE-Net.

  \subsection{Implementation Details}
  We implemented the proposed SGUIE-Net on PyTorch platform. During the training, the filter weights of each layer were initialized with Kaiming initialization. We used the Adam for network optimization and the initial learning rate was set to 1e-4 with the scheduler of linearly decaying to zero. Besides, batch size is set to 1. We make the input image size to $256 \times 256$ and perform data augmentation by random cropping and flipping. The experiments are conducted on a PC with an NVIDIA RTX 3090 GPU, a 2.3GHz Intel Xeon processor, 128GB RAM and Ubuntu 18.04 operation system.

\subsection{Experiment Settings}

\begin{figure*}[t]
  \begin{center}
  \includegraphics[width=1\linewidth]{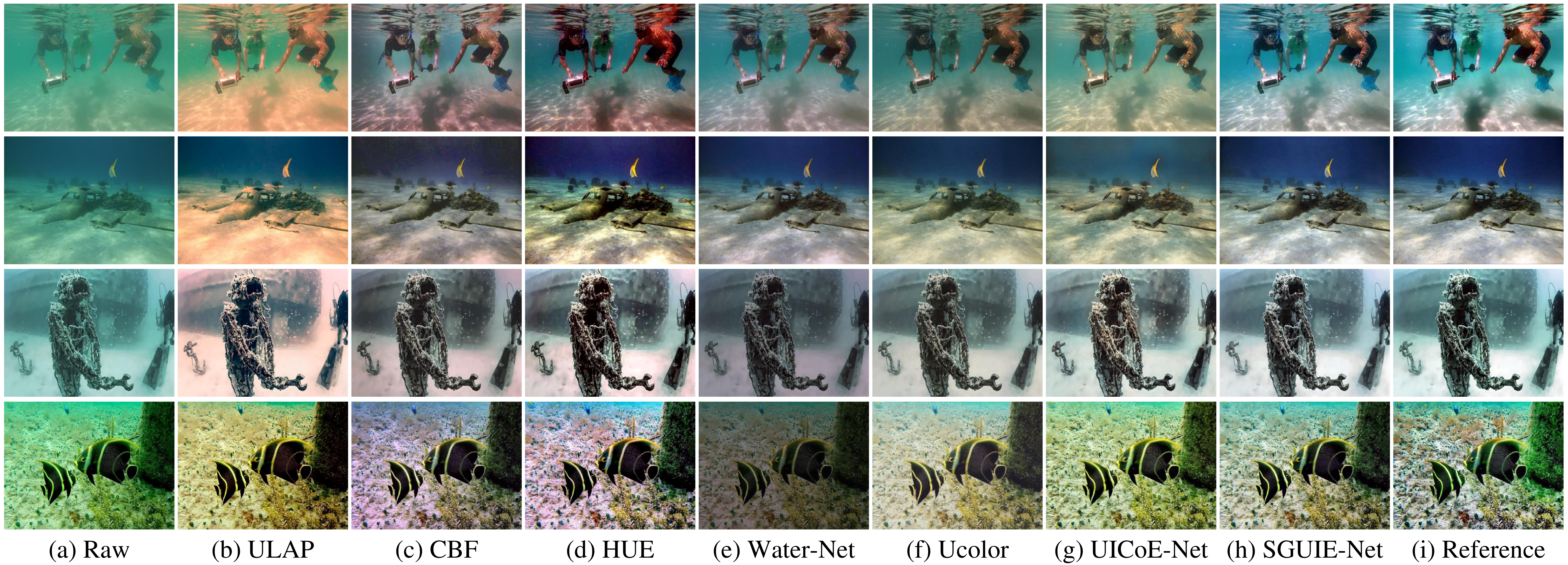}
  \caption{Visual comparisons on underwater images from SUIM-E test set. From left to right are raw underwater images and the results of ULAP\cite{Song2018ULAP}, CBF \cite{Ancuti2018Color}, HUE \cite{li2020hybrid}, Water-Net\cite{Li2020UIEBD}, Ucolor\cite{li2021underwater}, UICoE-Net\cite{Qi2021Underwater}, the proposed SGUIE-Net, and reference images.}
  \label{fig:SUIME}
  \end{center}
\end{figure*}

\subsubsection{Datasets}
As mentioned above, our approach uses semantic information as high-level guidance. To the best of our knowledge, there is no underwater dataset that contains both corresponding enhancement reference and semantic segmentation map. Therefore, to more fully verify the performance of our network, we constructed the SUIM-E dataset by supplementing the underwater images in SUIM dataset \cite{islam2020semantic} with corresponding enhancement references in a hand-picked manner similar to \cite{Li2020UIEBD}.
In detail, the SUIM-E dataset contains a total of 1635 real underwater images with pixel annotations for eight object categories: fish (vertebrates), reefs (invertebrates), aquatic plants, wrecks/ruins, human divers, robots, and sea-floor. We used 1525 of these images for training/validation and the remaining 110 images for testing.

We also conducted tests on the UIEB \cite{Li2020UIEBD} dataset, which includes 890 underwater images with manually selected enhancement references. We randomly selected 800 images for training/validation. The rest 90 images and another 60 challenging images without references are used for testing. Pre-trained SUIM-Net \cite{islam2020semantic} was adopted to predict semantic segmentation maps for both training and test.
In addition, we also verify the generalizability and color correction accuracy of our method on the RUIE \cite{liu2020real}, EUVP \cite{Islam2020EUVPFUnIEGAN}, SQUID \cite{berman2020underwater} and Color-Checker7 \cite{Ancuti2018Color} datasets with the model trained on  SUIM-E dataset. 

\subsubsection{Compared Methods}
We compared our SGUIE-Net with six methods proposed in recent 3 years, including one physical model-based method (ULAP\cite{Song2018ULAP}), two physical model-free methods (CBF \cite{Ancuti2018Color} and HUE \cite{li2020hybrid}), three deep learning-based methods (Water-Net\cite{Li2020UIEBD}, Ucolor\cite{li2021underwater}, UICoE-Net\cite{Qi2021Underwater}).
Since the source code of the CBF\cite{Ancuti2018Color} is not publicly available, as a compromise, we use the code\footnote{https://github.com/fergaletto/Color-Balance-and-fusion-for-underwater-image-enhancement.-.} reproduced by other researcher. Other comparison methods are used the released codes by their authors to produce their results.

\begin{table}[t]
  \centering
  \caption{Quantitative comparison on SUIM-E test set in terms of MSE$(\times{10}^{3})$, PSNR, SSIM, UIQM and UCIQE. Traditional methods and deep-learning-based methods trained with paired reference images are separated with lines. The top three scores are marked in red, green and blue.}
  \label{tab:SUIME}
  \fontsize{9pt}{11pt}\selectfont
  \begin{tabular}{c||C{0.08\linewidth} C{0.09\linewidth} C{0.09\linewidth} C{0.09\linewidth} C{0.11\linewidth}}
  \Xhline{1pt}
  Method & MSE$\downarrow$ & PSNR$\uparrow$ & SSIM$\uparrow$ & UIQM$\uparrow$ & UCIQE$\uparrow$ \\
  \hline \hline
  ULAP\cite{Song2018ULAP} & 1.752 & 16.561 & 0.769 & 0.616 & \textcolor{blue}{0.614} \\
  CBF\cite{Ancuti2018Color} & 1.804 & 16.465 & 0.828 & \textcolor{green}{0.836} & 0.571 \\
  HUE\cite{li2020hybrid} & 1.268 & 17.97 & 0.822 & \textcolor{red}{1.022} & \textcolor{red}{0.651} \\
  \hline           
  Water-Net\cite{Li2020UIEBD} & 1.628 & 17.019 & 0.82 & 0.415 & 0.548 \\
  Ucolor\cite{li2021underwater} & \textcolor{blue}{0.639} & \textcolor{blue}{21.063} & \textcolor{blue}{0.844} & 0.5 & 0.577 \\
  UICoE-Net\cite{Qi2021Underwater} & \textcolor{green}{0.550} & \textcolor{green}{21.752} & \textcolor{green}{0.910} & 0.668 & 0.588 \\
  SGUIE-Net & \textcolor{red}{0.307} & \textcolor{red}{24.820} & \textcolor{red}{0.928} & \textcolor{blue}{0.703} & \textcolor{green}{0.615} \\
  \hline
  \Xhline{0.2pt}
  \end{tabular}
\end{table}

\begin{figure}[t]
  \begin{center}
  \includegraphics[width=1\linewidth]{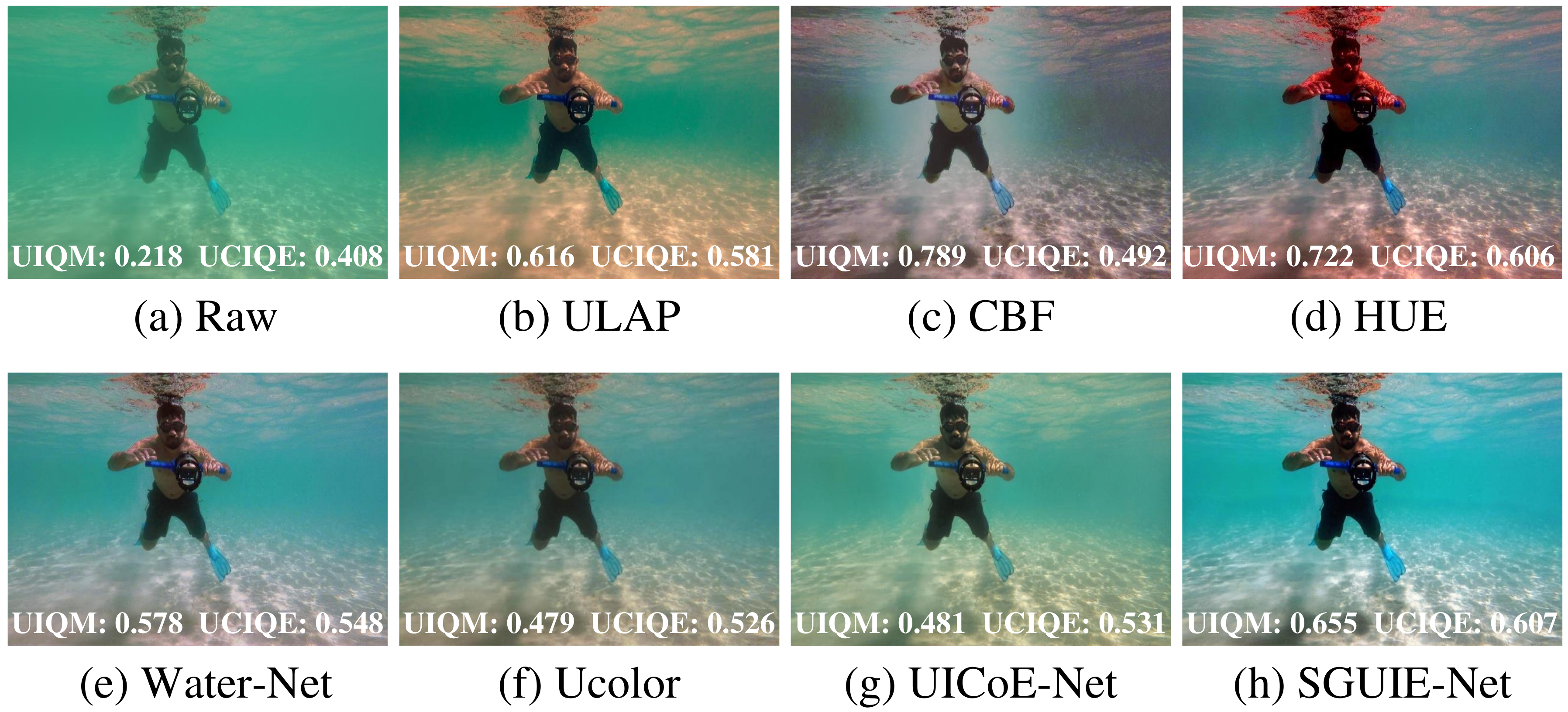}
  \caption{Scores of no-reference quality metrics on enhancements of an underwater image from SUIM-E test set. (a)-(h) are raw underwater images and the results of ULAP\cite{Song2018ULAP}, CBF \cite{Ancuti2018Color}, HUE \cite{li2020hybrid}, Water-Net\cite{Li2020UIEBD}, Ucolor\cite{li2021underwater}, UICoE-Net\cite{Qi2021Underwater} and the proposed SGUIE-Net. The UIQM and UCIQE scores are marked at the bottom.}
  \label{fig:UIQM_UCIQE}
  \end{center}
\end{figure}

\begin{figure*}[t]
  \begin{center}
  \includegraphics[width=1\linewidth]{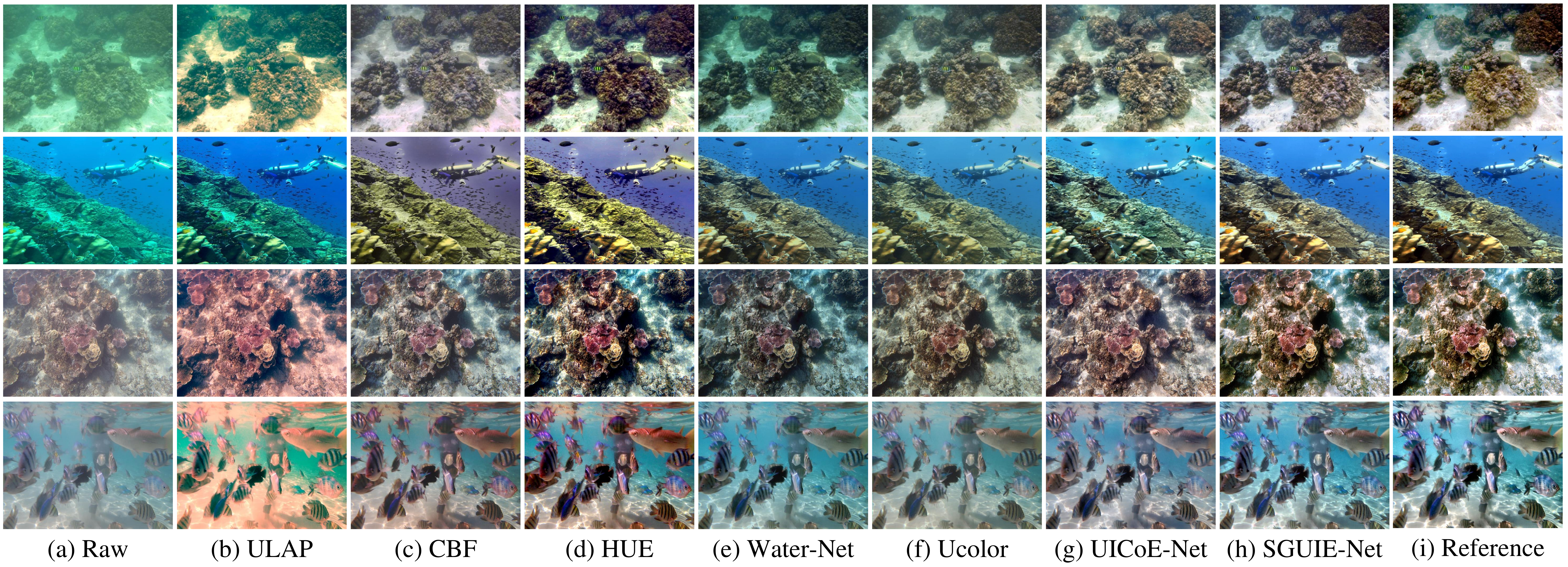}
  \caption{Visual comparisons on underwater images from UIEB test set. From left to right are raw underwater images and the results of ULAP\cite{Song2018ULAP}, CBF \cite{Ancuti2018Color}, HUE \cite{li2020hybrid}, Water-Net\cite{Li2020UIEBD}, Ucolor\cite{li2021underwater}, UICoE-Net\cite{Qi2021Underwater}, the proposed SGUIE-Net, and reference images.}
  \label{fig:UIEB}
  \end{center}
\end{figure*}

\subsection{Evaluation of Visual Enhancement Performance}

In this section, we perform qualitative and quantitative analyses on the SUIM-E, UIEB, RUIE and EUVP datasets. Figure \ref{fig:SUIME} shows the results of different comparison methods on the SUIM-E dataset.
The traditional methods, such as ULAP \cite{Song2018ULAP}, CBF\cite{Ancuti2018Color} and HUE \cite{li2020hybrid}, usually cause color deviation due to the introduction of excessive red components. 
Since the Water-Net method introduces a white balance channel during the enhancement process, which is not always reliable for underwater images, this may leads to color bias as well. 
The enhancement results of the Ucolor\cite{li2021underwater} achieve a better visual quality from the perspective of color correction. However, as shown in the first three rows of the examples, Ucolor does not improve the low contrast and blurred details very well. 
UICoE-Net \cite{Qi2021Underwater} requires paired images matching similar scenes for correlation feature learning. Therefore, give SUIM-E dataset without any scene classification, it may lead to unstable performance of the enhancement of the UICoE-Net method.
On the contrary, our SGUIE-Net provides stable color correction in severe color deviation environments and does not introduce extra colors deviation. Our method is able to naturally improve the contrast while enhancing it in low-light underwater scenes.
In particular, as shown in the third row of the Figure \ref{fig:SUIME}, the enhancement of the white sands by our method is more complete compared to other comparison methods, and the visual quality even outperforms the reference enhanced image.
The comparison shows that our method achieves a better balance between visual quality and color correction. Thanks to the cascade attention-aware enhancement module, our enhancement results can preserve the texture details of the original image to the maximum extent. Moreover, with the high-level semantic guidance, our approach can apply consistent enhancements to the same semantic regions, such as the white-sand sea-floor in the top three rows. Besides, it is also able to fine-tune the results according to the scene context through semantic region-wise enhancement module, resulting in more natural results.

\begin{table}[t]
  \centering
  \caption{Quantitative comparison on UIEB test set in terms of MSE$(\times{10}^{3})$, PSNR, SSIM, UIQM and UCIQE. Traditional methods and deep-learning-based methods trained with paired reference images are separated with lines. The top three scores are marked in red, green and blue.}
  \label{tab:UIEB}
  \fontsize{9pt}{11pt}\selectfont
  \begin{tabular}{c||C{0.08\linewidth} C{0.09\linewidth} C{0.09\linewidth} C{0.09\linewidth} C{0.11\linewidth}}
  \Xhline{1pt}
  Method & MSE$\downarrow$ & PSNR$\uparrow$ & SSIM$\uparrow$ & UIQM$\uparrow$ & UCIQE$\uparrow$ \\
  \hline \hline
  ULAP\cite{Song2018ULAP} & 1.929 & 16.786 & 0.776 & 0.742 & \textcolor{blue}{0.598} \\
  CBF\cite{Ancuti2018Color} & 1.419 & 17.678 & 0.849 & \textcolor{green}{1.028} & 0.551 \\
  HUE\cite{li2020hybrid} & 1.172 & 18.63 & 0.834 & \textcolor{red}{1.260} & \textcolor{red}{0.648} \\
  \hline
  Water-Net\cite{Li2020UIEBD} & 0.955 & 19.298 & \textcolor{blue}{0.873} & 0.501 & 0.563 \\
  Ucolor\cite{li2021underwater} & \textcolor{green}{0.693} & \textcolor{green}{20.742} & 0.847 & 0.802 & 0.548 \\
  UICoE-Net\cite{Qi2021Underwater} & \textcolor{blue}{0.800} & \textcolor{blue}{20.070} & \textcolor{green}{0.875} & 0.838 & 0.585 \\
                        
  SGUIE-Net & \textcolor{red}{0.381} & \textcolor{red}{24.074} & \textcolor{red}{0.908} & \textcolor{blue}{0.851} & \textcolor{green}{0.601} \\
  \hline
  \Xhline{0.2pt}
  \end{tabular}
\end{table}

The quantitative numerical results for the SUIM-E dataset are shown in Table \ref{tab:SUIME}. We used the full-reference evaluation metrics including structural similarity index metric (SSIM), peak signal-to-noise ratio (PSNR) and mean square error (MSE) as the main evaluation metrics, supplemented with the no-reference evaluation metrics including underwater image quality measure (UIQM) \cite{Panetta2016UIQM} and underwater color image quality evaluation (UCIQE)\cite{Yang2015UCIQE} metrics. 
By comparing the results of all metrics in combination, we can observe that our method achieves better results overall. Specifically, our method improves 14.1\% and 44.2\% in PSNR and MSE metrics compared to the second-best method. For the SSIM metric, UICoE-Net scores 0.910 since it introduces correlation feature matching units for better local detail recovery. While, our SGUIE-Net achieved a better SSIM score of 0.928, which further demonstrates the superiority of our method in maintaining detailed image texture.
In addition, Our method scores in the top three for both UIQM and UCIQE metrics, and ranks first among deep learning methods in the comparison. Since the HUE method contains operations for color saturation and contrast enhancement, the CBF contains processes for white balance as well as edge sharpening, which are preferred by UIQM and UCIQE. Thus, it also leads to the fact that these two methods can achieve relatively higher scores in that two metrics. However, as shown in Figure \ref{fig:UIQM_UCIQE}, the visual quality of HUE and CBF is not satisfactory in many underwater scenes, and this is reflected in the evaluation of full-reference metrics.

To further verify the effectiveness and robustness of our method, we conducted experiments on UIEB dataset. The visual comparisons with other methods are shown in the Figure \ref{fig:UIEB}. The underwater images shown in Figure \ref{fig:UIEB} suffer from color deviation while also experiencing severe backscatter, which raise great challenges for the enhancement methods.
The ULAP, CBF and HUE either fail to enhance the input image or introduce the over-saturation or artifacts. 
By looking at the first two rows of Figure \ref{fig:UIEB}, most comparison methods are not suitable for accurate color correction of severely degraded input images containing large amounts of green veils.
In addition, it can be seen that the severe backscatter affects the performance of Ucolor and Water-Net, making the results less clear and  the deeper regions in the underwater image not effectively enhanced. 
As a comparison, our method can handle enhancement properly in scenes with severe backscatter and is able to perform color correction incorporating semantic guidance while maintaining the details of the input image.

\begin{figure*}[t]
  \begin{center}
  \includegraphics[width=1\linewidth]{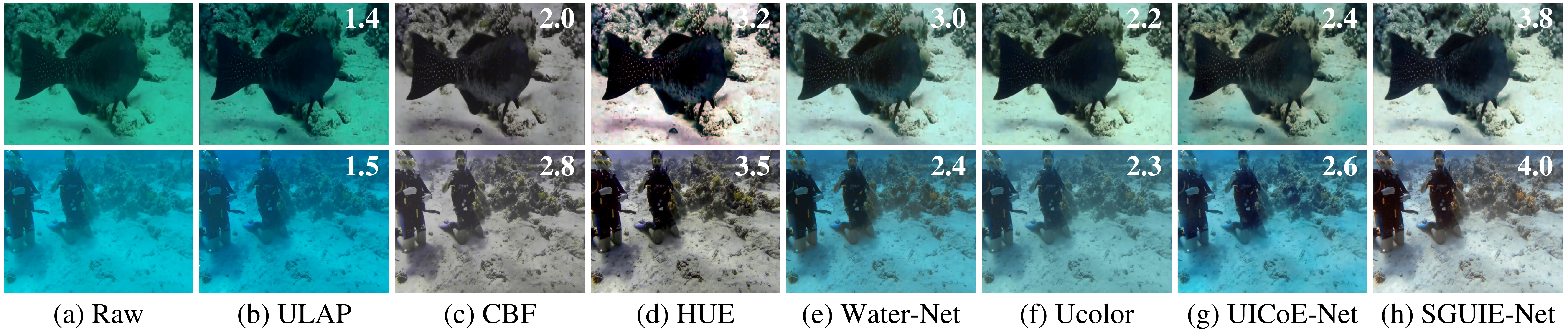}
  \caption{Visual comparisons on underwater images from UIEB challenging test set. From left to right are raw underwater images and the results of ULAP\cite{Song2018ULAP}, CBF \cite{Ancuti2018Color}, HUE \cite{li2020hybrid}, Water-Net\cite{Li2020UIEBD}, Ucolor\cite{li2021underwater}, UICoE-Net\cite{Qi2021Underwater} and the proposed SGUIE-Net. Perceptual scores are marked on the upper right corner.}
  \label{fig:UIEB-challenging}
  \end{center}
\end{figure*}

\begin{figure*}[t]
  \begin{center}
  \includegraphics[width=1\linewidth]{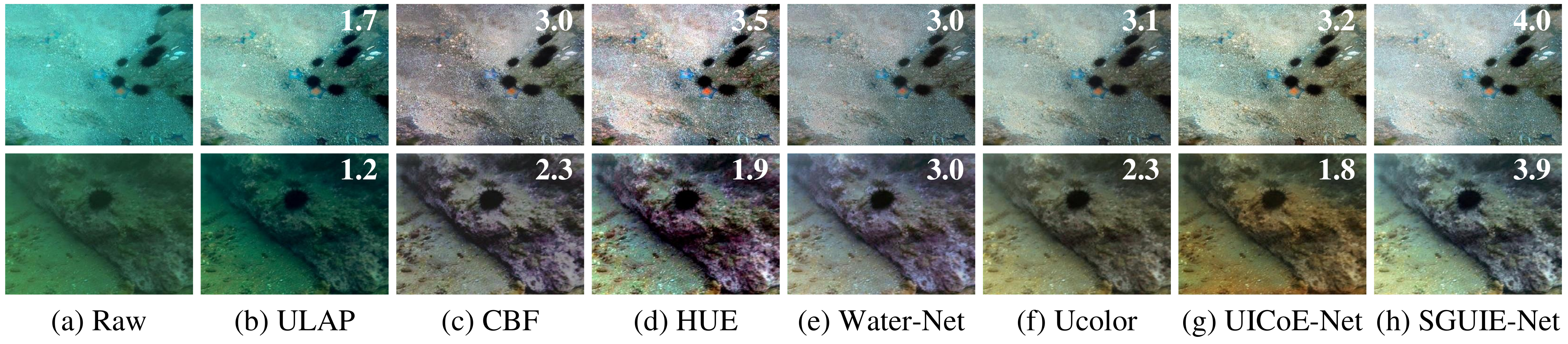}
  \caption{Visual comparisons on underwater images from RUIE dataset. From left to right are raw underwater images and the results of ULAP\cite{Song2018ULAP}, CBF \cite{Ancuti2018Color}, HUE \cite{li2020hybrid}, Water-Net\cite{Li2020UIEBD}, Ucolor\cite{li2021underwater}, UICoE-Net\cite{Qi2021Underwater} and the proposed SGUIE-Net. Perceptual scores are marked on the upper right corner.}
  \label{fig:RUIE}
  \end{center}
\end{figure*}

\begin{figure*}[t]
  \begin{center}
  \includegraphics[width=1\linewidth]{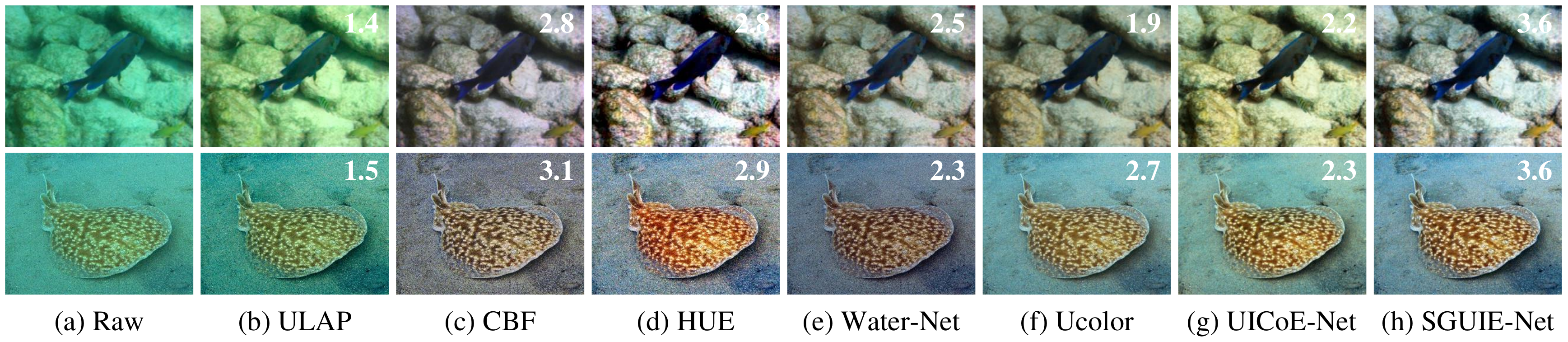}
  \caption{Visual comparisons on underwater images from EUVP dataset. From left to right are raw underwater images and the results of ULAP\cite{Song2018ULAP}, CBF \cite{Ancuti2018Color}, HUE \cite{li2020hybrid}, Water-Net\cite{Li2020UIEBD}, Ucolor\cite{li2021underwater}, UICoE-Net\cite{Qi2021Underwater} and the proposed SGUIE-Net. Perceptual scores are marked on the upper right corner.}
  \label{fig:EUVP}
  \end{center}
\end{figure*}

\begin{table*}[!htb]
  \centering
  \caption{No-reference image quality assessment in terms of UIQM, UCIQE and Average Perceptual Scores (PS) on UIEB challenging set, RUIE dataset and EUVP dataset. Traditional methods and deep-learning-based methods trained with paired reference images are separated with lines. The top three scores are marked in red, green and blue.}
  \label{tab:UIEB_RUIE_EUVP}
  \fontsize{9pt}{11pt}\selectfont
  \begin{tabular}{c||C{0.06\linewidth} C{0.06\linewidth} C{0.06\linewidth} | C{0.06\linewidth} C{0.06\linewidth} C{0.06\linewidth} | C{0.06\linewidth} C{0.06\linewidth} C{0.06\linewidth}}
  \Xhline{1pt}
  \multirow{2}{*}{Method} & \multicolumn{3}{c|}{UIEB Challenging} & \multicolumn{3}{c|}{RUIE} & \multicolumn{3}{c}{EUVP} \\
  \cline{2-10}
   & PS$\uparrow$ & UIQM$\uparrow$ & UCIQE$\uparrow$ & PS$\uparrow$ & UIQM$\uparrow$ & UCIQE$\uparrow$ & PS$\uparrow$ & UIQM$\uparrow$ & UCIQE$\uparrow$ \\
  \hline \hline
  ULAP\cite{Song2018ULAP} & 1.768 & 0.368 & 0.540 & 1.349 & 0.162  & 0.490  & 1.878 & 0.954  & \textcolor{blue}{0.600} \\
  CBF\cite{Ancuti2018Color} & 2.440 & \textcolor{green}{0.802} & 0.508 & \textcolor{green}{2.681} & \textcolor{green}{0.773}  & 0.489  & \textcolor{blue}{2.498} & \textcolor{green}{1.156}  & 0.562 \\
  HUE\cite{li2020hybrid} & 2.503 & \textcolor{red}{0.873} & \textcolor{red}{0.600} & 2.336 & \textcolor{red}{1.078}  & \textcolor{red}{0.608}  & 2.436 & \textcolor{red}{1.443}  & \textcolor{red}{0.638} \\
  \hline
  Water-Net\cite{Li2020UIEBD} & 2.623 & 0.248 & \textcolor{blue}{0.549} & 2.588 & 0.590  & \textcolor{blue}{0.526}  & 1.958 & 0.709  & 0.534 \\
  Ucolor\cite{li2021underwater} & \textcolor{green}{2.658} & 0.265 & 0.520 & \textcolor{blue}{2.589} & 0.621  & \textcolor{blue}{0.526}  & 2.488 & 0.814  & 0.561 \\
  UICoE-Net\cite{Qi2021Underwater} & \textcolor{blue}{2.642} & 0.312 & 0.539 & 2.092 & 0.517  & 0.518  & \textcolor{green}{2.681} & 0.910  & 0.586 \\ 
  SGUIE-Net & \textcolor{red}{3.283} & \textcolor{blue}{0.524} & \textcolor{green}{0.578} & \textcolor{red}{3.529} & \textcolor{blue}{0.688}  & \textcolor{green}{0.556}  & \textcolor{red}{3.308} & \textcolor{blue}{1.007} & \textcolor{green}{0.603} \\
  \hline
  \Xhline{0.2pt}
  \end{tabular}
\end{table*}

\begin{figure*}[!ht]
  \begin{center}
  \includegraphics[width=1\linewidth]{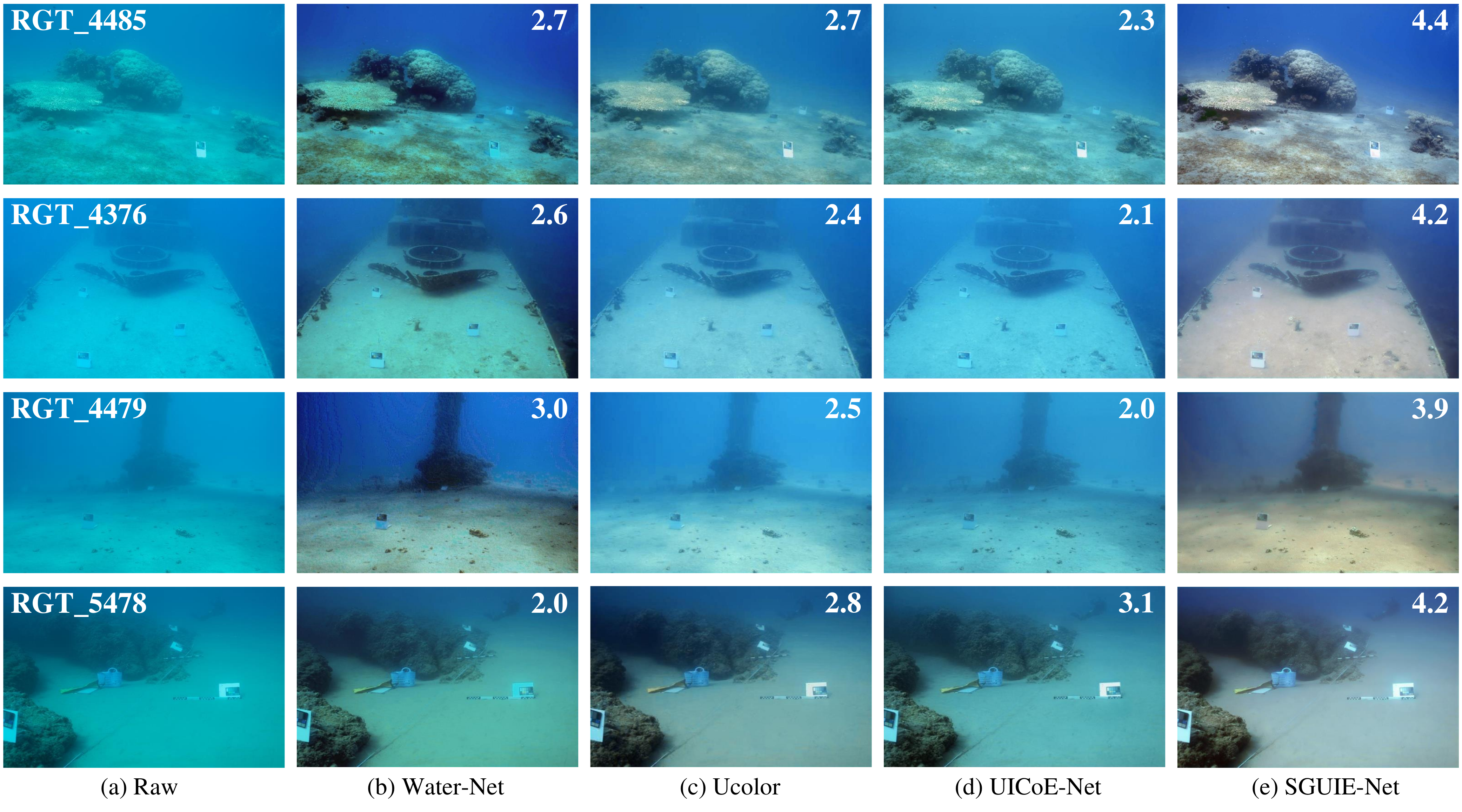}
  \caption{Visual comparisons on challenging underwater images from SQUID. From left to right are raw underwater images, and the results of Water-Net\cite{Li2020UIEBD}, Ucolor\cite{li2021underwater}, UICoE-Net\cite{Qi2021Underwater} and the proposed SGUIE-Net. Perceptual scores are marked on the upper right corner.}
  \label{fig:SQUID}
  \end{center}
  \vspace{-5mm}
\end{figure*}

\begin{figure*}[!htb]
  \centering
  \subfigure
  {
      \begin{minipage}[b]{.49\linewidth}
          \centering
          \includegraphics[width=3.in, height=1.46in]{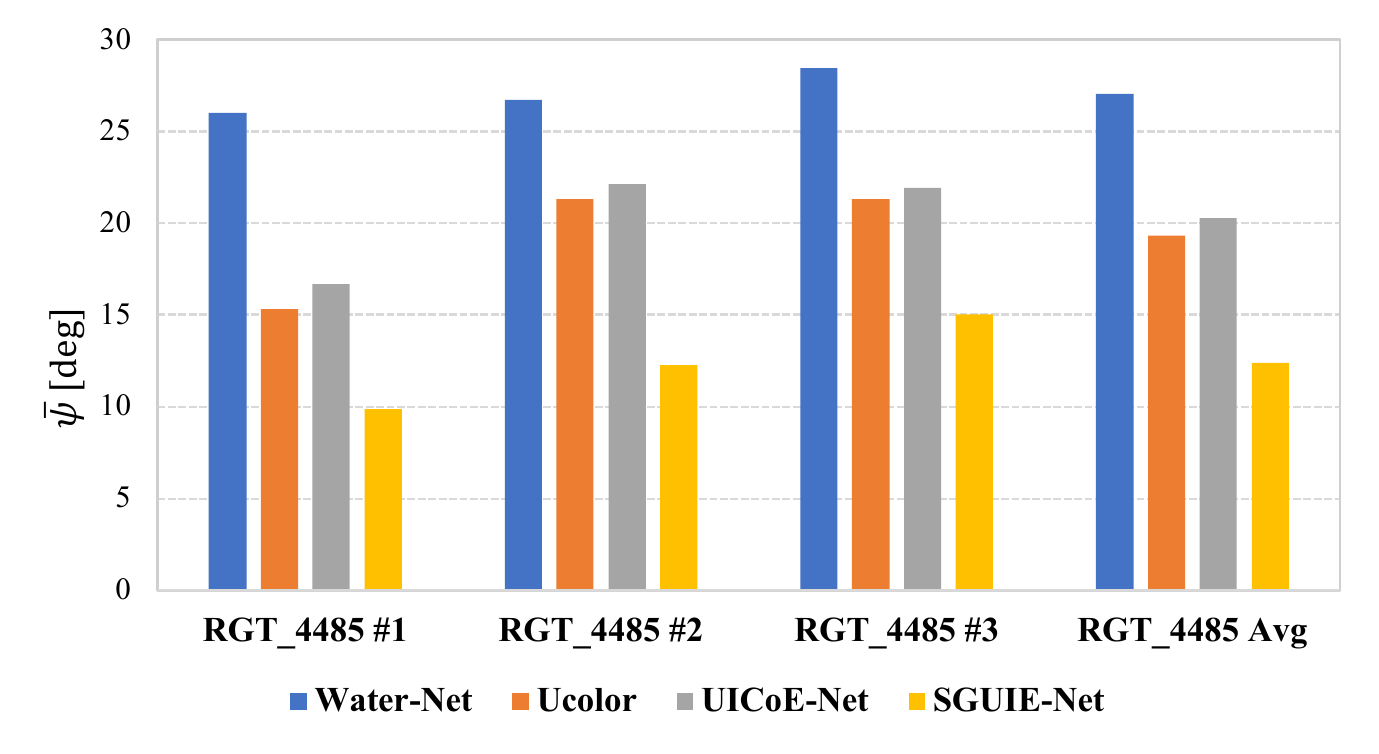}
      \end{minipage}
  }\hspace{-3mm}
  \subfigure
  {
      \begin{minipage}[b]{.49\linewidth}
          \centering
          \includegraphics[width=3.in, height=1.46in]{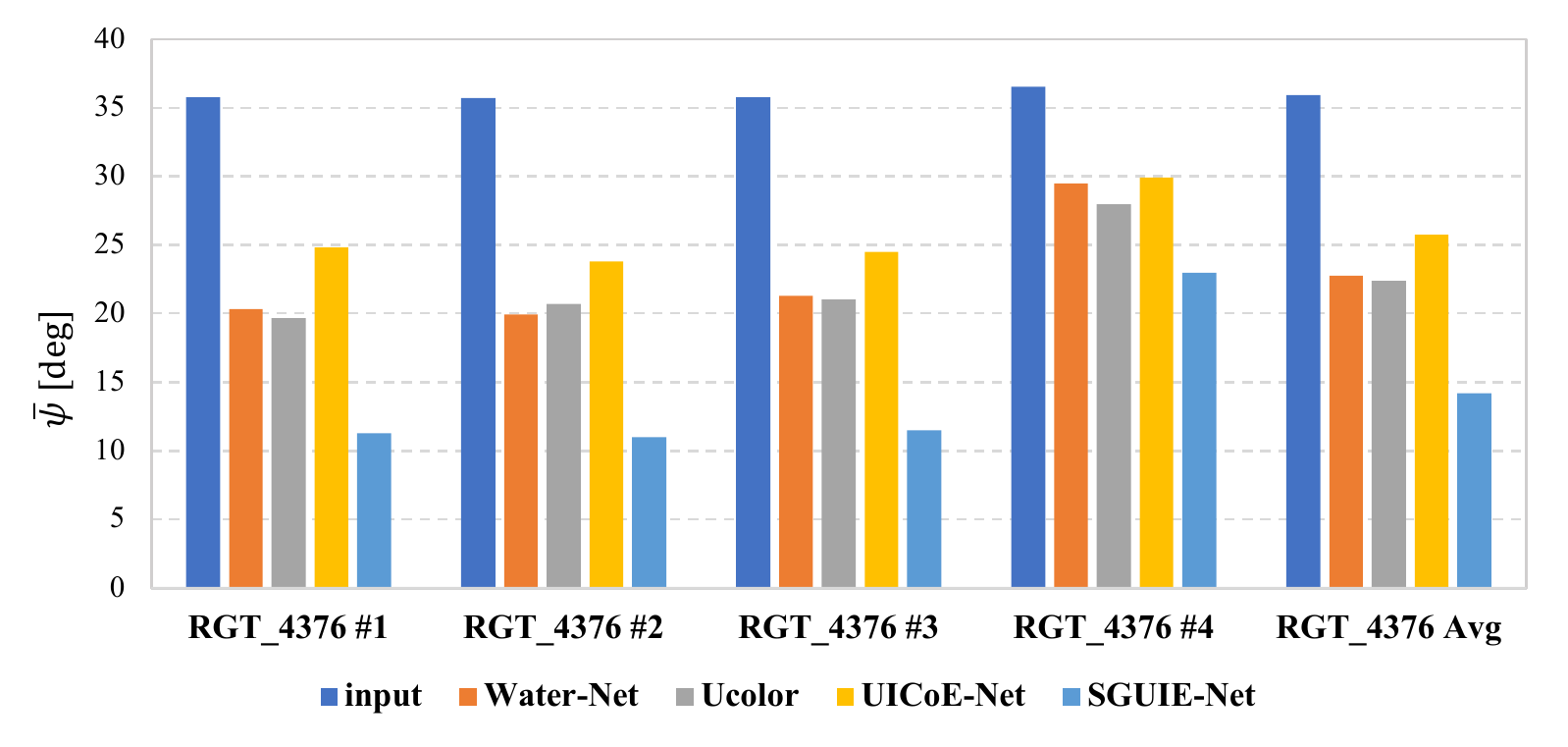}
      \end{minipage}
  }\vspace{-4mm}

  \subfigure
  {
     \begin{minipage}[b]{.49\linewidth}
          \centering
          \includegraphics[width=3.in, height=1.46in]{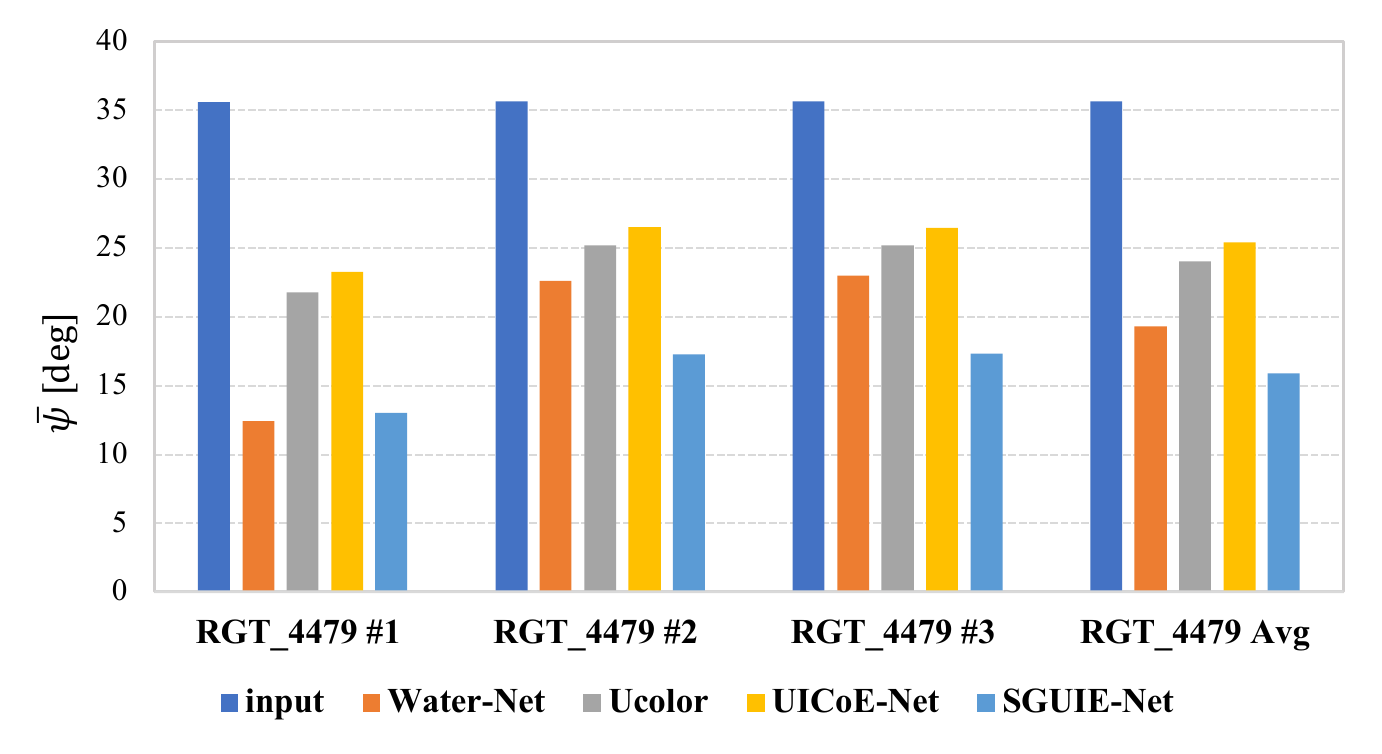}
      \end{minipage}
  }\hspace{-2.8mm}
  \subfigure
  {
     \begin{minipage}[b]{.49\linewidth}
          \centering
          \includegraphics[width=3.in, height=1.46in]{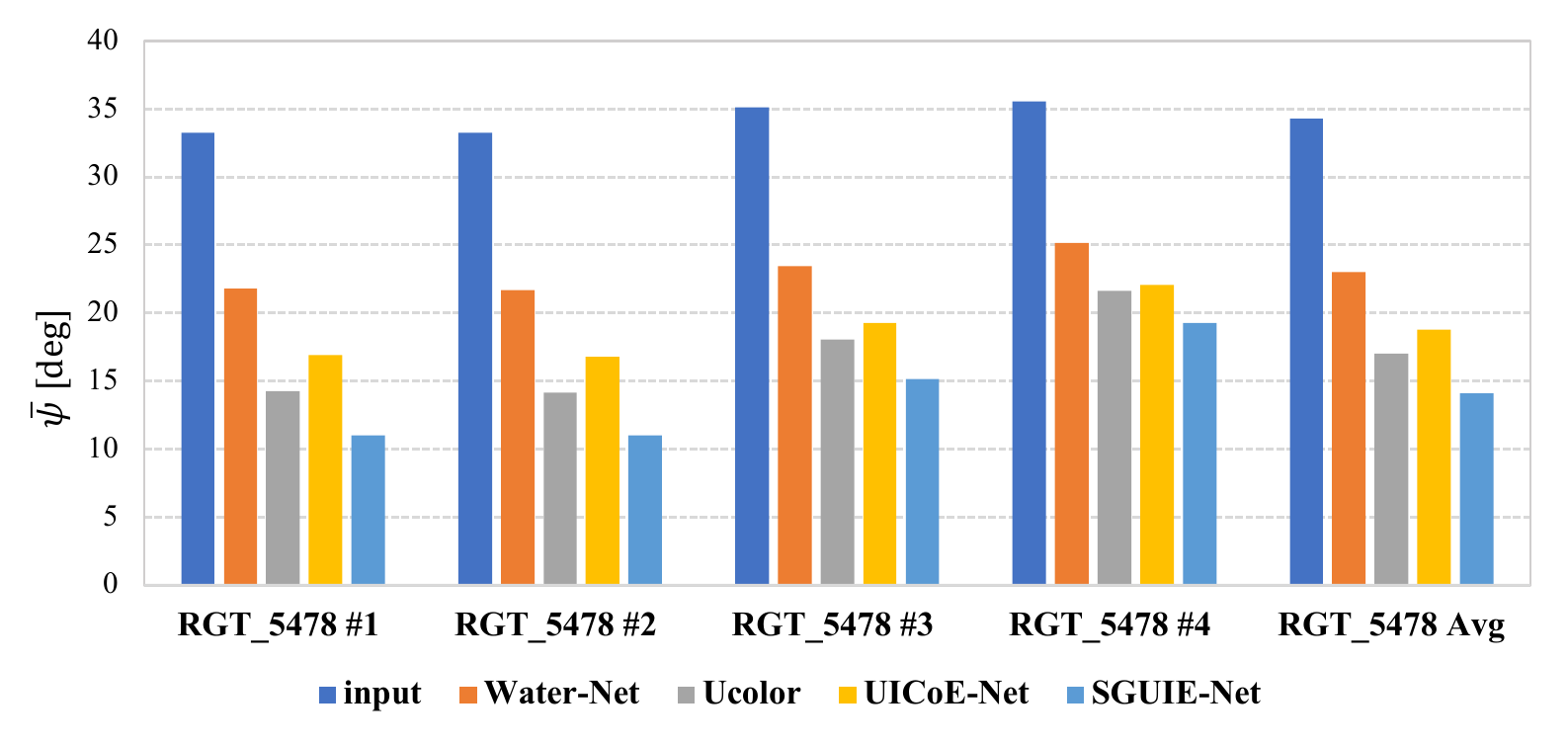}
      \end{minipage}
  }

  \caption{Comparison of average reproduction angular error $ \overline{\psi}$ between the gray-scale patches and the pure gray colors. The sub-bargraph reports the average performance of each comparison deep learning method on all color charts in test image. The average performance of each approach is reported on the right.}
  \label{fig:SQUID_h}
  \vspace{-3mm}
  \end{figure*}

  \begin{table}[!ht]
    \centering
    \caption{The Average angular reproduction error (AE), PS, UIQM and UCIQE on SQUID. The best score is in red under each metric.}
    \label{tab:SQUID}
    \fontsize{9pt}{11pt}\selectfont
    \begin{tabular}{c||c c c c}
    \Xhline{1pt}
    Method & AE$\downarrow$ & PS$\uparrow$ & UIQM$\uparrow$ & UCIQE$\uparrow$ \\
    \hline \hline
    Water-Net\cite{Li2020UIEBD} & 20.105  & 2.288 & 0.124 & 0.528 \\
    Ucolor\cite{li2021underwater} & 20.771  & 2.683 & 0.153 & 0.498 \\
    UICoE-Net\cite{Qi2021Underwater} & 20.788  & 2.619 & 0.254 & 0.482 \\
    SGUIE-Net & \textcolor{red}{14.675} & \textcolor{red}{3.793} & \textcolor{red}{0.271} & \textcolor{red}{0.553} \\
    \hline
    \Xhline{0.2pt}
    \end{tabular}
  \end{table}

  \begin{table*}[t]
    \centering
    \caption{The color dissimilarity comparisons of different methods on color-check7 in terms of the ciede2000. Traditional methods and deep learning-based methods trained with paired reference images are separated with lines. The best scores are marked in red.}
    \label{tab:ColorCheck7}
    \fontsize{9pt}{11pt}\selectfont
    \begin{tabular}{c||C{0.08\linewidth} C{0.09\linewidth} C{0.08\linewidth} C{0.08\linewidth} C{0.08\linewidth} C{0.08\linewidth} C{0.08\linewidth} C{0.08\linewidth}}
    \Xhline{1pt}
    Method & Can D10 & Fuji Z33 & Oly T6000 & Oly T8000 & Pan TS1 & Pen W60 & Pen W80 & Avg \\
    \hline \hline
    Input &12.910  &16.648  &14.990  & 19.301  & 16.152  & 11.966  & 14.123  & 15.156 \\
    ULAP\cite{Song2018ULAP} & 16.788  & 11.671  & 11.508  & 17.176  & 23.469  & 17.729  & 18.547  & 16.698 \\
    CBF\cite{Ancuti2018Color} & \textcolor{red}{9.447} & 10.249 & 9.656 & 12.746 & 9.815 & \textcolor{red}{9.664} & 11.458 & 10.434 \\
    HUE\cite{li2020hybrid} & 14.214 & 12.863 & 12.943 & 13.785 & 12.496 & 14.287 & 14.003 & 13.513 \\
    \hline
    Water-Net\cite{Li2020UIEBD} & 13.920 & 18.779 & 12.132 & 16.889 & 18.971 & 11.851 & 17.630 & 15.739 \\
    Ucolor\cite{li2021underwater} & 10.424 & 10.706 & 8.758 & 11.668 & 10.587 & 10.110 & 10.678 & 10.419 \\
    UICoE-Net\cite{Qi2021Underwater} & 10.023 & 13.085 & 13.291 & 11.846 & 11.438 & 10.416 & \textcolor{red}{10.159} & 11.465 \\
    
    SGUIE-Net & 10.526 & \textcolor{red}{10.227} & \textcolor{red}{8.654} & \textcolor{red}{11.292} & \textcolor{red}{9.572} & 10.999 & 11.504 & \textcolor{red}{10.396} \\
    \hline
    \Xhline{0.2pt}
    \end{tabular}
  \end{table*}
  
  \begin{figure*}[t]
    \begin{center}
    \includegraphics[width=1\linewidth]{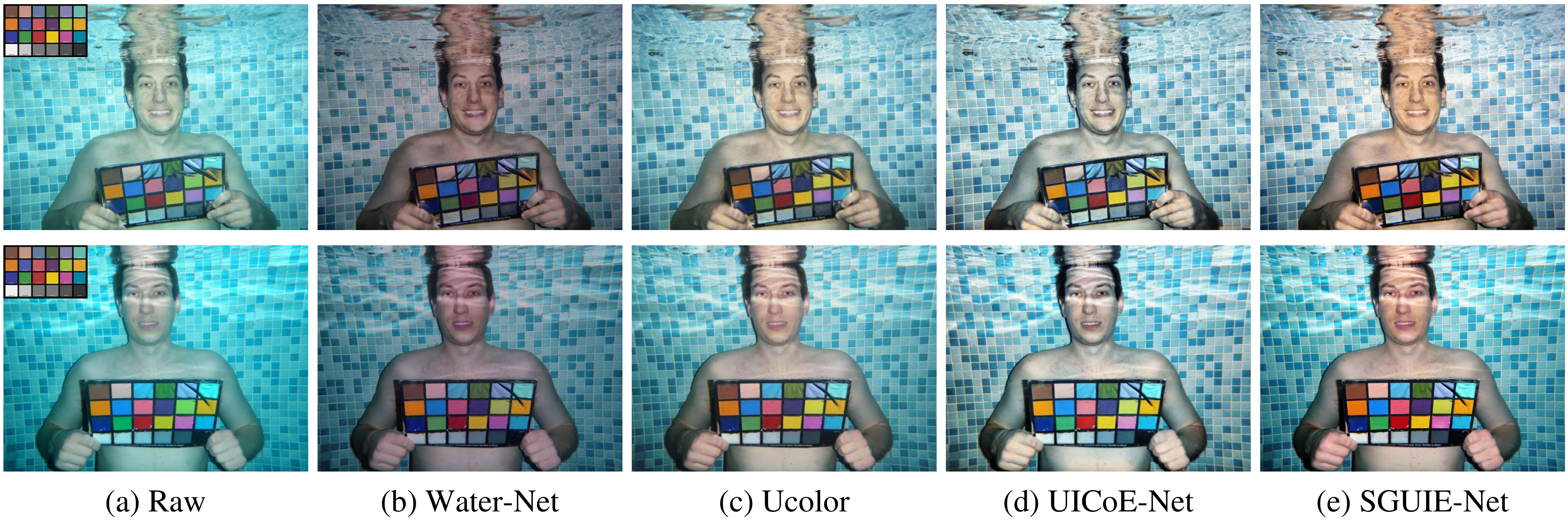}
    \caption{Visual comparisons on images of Color-Checker7. The two images are taken by the Fuji Z33 (first row) and Olympus T6000 (second row), respectively. From left to right are raw underwater images and the results of Water-Net\cite{Li2020UIEBD}, Ucolor\cite{li2021underwater}, UICoE-Net\cite{Qi2021Underwater} and the proposed SGUIE-Net.}
    \label{fig:ColorCheck7}
    \end{center}
  \end{figure*}

The numerical comparison on UIEB dataset is given in Table \ref{tab:UIEB}. Among all the underwater image enhancement methods we test, our SGUIE-Net achieves the best results on all three full-reference metrics evaluations, which is consistent with the qualitative analysis given above. By comparison, data-driven-based methods generally achieve higher scores than traditional methods, which is a further evidence of the fact that the traditional physical-based methods cannot well adapt to varied underwater environments. Comparing the scores of our method on the SUIM-E and the UIEB datasets, we can find that the test scores of SGUIE-Net on the UIEB test set are lower compared to those on the SUIM-E test set. In fact, this is can be anticipated in advance since the ground truth semantic guide information is not available on the UIEB dataset. Therefore, we used the predicted semantic segmentation results generated by the SUIM-Net as a substitute. Note that we directly use the pre-trained model without any fine-tune, which leads to the possibility that some of the semantic information obtained by our network is not exactly correct, making our network unable to achieve the best performance. However, even with the incomplete semantic information, our method achieves the best score among all the compared methods. 

To demonstrate the performance and robustness of our method on tough cases, we also conducted comparative tests on the challenging set of UIEB. As shown in Figure \ref{fig:UIEB-challenging}, these underwater images suffer from low-lighting and color deviation. ULAP failed to enhanced these challenging images.
The results of HUE introduce excessive color saturation and contrast, making the results appear unrealistic. Although CBF removes extra green component, the overall contrast of the results is still low. Water-Net and UICoE-Net fail in extreme scenarios that have not been well-learned before, such as the example shown in the second row. Ucolor improve the visual quality but could not achieve satisfactory enhancement.
In contrast, our SGUIE-Net is able to handle such low-light and color deviation environment well, and can smoothly boost image contrast, making the results more vivid and natural.

We also conducted comprehensive test and evaluation on the RUIE and EUVP datasets to evaluate the generalization and robustness of our model. 
As shown in Figure \ref{fig:RUIE} and Figure \ref{fig:EUVP}, CBF and HUE introduce extra colors while enhancing the images, which affects the quality of the results. Water-Net, UICoE-Net and Ucolor are effective in removing bluish and greenish distortion, but the results are not satisfactory in terms of contrast, brightness and detailed textures. Based on the already constructed semantic-related mappings between the degradation and enhancement, our SGUIE-Net can apply adaptive enhancements to different semantic regions with such high-level guidance. Especially for raw images that have suffered severe degradation and whose details appear blurred, our method can also enhance texture details while performing color correction, resulting in more pleasant results.

Since there are no corresponding enhancement references for the images of these three datasets, we perform user studies to give perceptual scores for more objective quantitative comparisons. 
Inspired by the Ucolor \cite{li2021underwater}, we invited 10 volunteers with basic knowledge of image processing to independently score the perceptual quality of the enhanced images. The perceptual quality was scored on a scale from 1 to 5 (from the worst to the best quality).
Considering the expensive labor costs and the fact that there are thousands of test images, we only select the representative portion of images for the evaluation of perception scores. Specifically, we sampled 10 images proportionally from each subfolder of the RUIE dataset, resulting in 130 images. In addition, we take the first 30 images from each subfolder of the EUVP dataset separately, resulting in 120 images.

The average perception scores for the results of different methods on these three datasets are reported in Table \ref{tab:UIEB_RUIE_EUVP}. Most deep learning-based methods outperform traditional methods in terms of visual perception quality metrics. Benefiting from the guidance of semantic high-level information, our method is able to maintain a relatively stable enhancement performance for challenging underwater images and thus achieves the highest visual perceptual quality score. In addition, our method also achieves the highest UIQM and UCIQE scores when comparing with the deep learning-based methods.

For more visual comparison with perception scores on images from UIEB challenging test set, RUIE and EUVP datasets, please refer to the supplement materials.

\subsection{Evaluation of Color Restoration Performance}

To analyse the robustness and accuracy of color restoration, we further conduct the comparisons on SQUID and Color-Checker7 datasets. For all the comparison methods, we use their models trained on the SUIM-E dataset for test.

\begin{table*}[!ht]
  \centering
  \caption{Quantitative results of the network modules ablation study on SUIM-E and UIEB test sets.}
  \label{tab:Eva_AB1}
  \fontsize{9pt}{11pt}\selectfont
  \begin{tabular}{C{0.1\linewidth}|| C{0.1\linewidth} | C{0.11\linewidth} C{0.08\linewidth} C{0.08\linewidth} | C{0.11\linewidth} C{0.08\linewidth} C{0.08\linewidth}}
  \Xhline{1pt}
  \multirow{2}{*}{Modules} & \multirow{2}{*}{Baselines} & \multicolumn{3}{c|}{SUIM-E} & \multicolumn{3}{c}{UIEB} \\
  \cline{3-8}
   & & MSE $(\times{10}^{3})\downarrow$ & PSNR$\uparrow$ & SSIM$\uparrow$ & MSE $(\times{10}^{3})\downarrow$ & PSNR$\uparrow$ & SSIM$\uparrow$ \\
  \hline \hline 
  CAM & w/o CAM & 0.497 & 22.000 & 0.893 & 0.598 & 21.200 & 0.863 \\
  \hline
  \multirow{3}{*}{SRM} & w/o SRM & 0.504 & 21.940 & 0.897 & 0.796 & 19.910 & 0.852 \\
   & w/o SS & 0.418 & 22.841 & 0.907 & 0.598 & 21.175 & 0.874 \\
   & w/o UNet & 0.446 & 22.587 & 0.908 & 0.579 & 21.357 & 0.877 \\
   \hline
  \multirow{3}{*}{FAB} & w/o FAB & 0.412 & 22.887 & 0.902 & 0.682 & 20.524 & 0.865 \\
  & w/o CA & 0.366 & 23.175 & 0.902 & 0.530 & 21.824 & 0.871 \\ 
   & w/o PA & 0.410 & 23.356 & 0.914 & 0.493 & 22.164 & 0.885 \\
   
  \hline
  full & --- & 0.307 & 24.820 & 0.928 & 0.381 & 24.074 & 0.908 \\
  \hline
  \Xhline{0.2pt}
  \end{tabular}
\end{table*}

In Figure \ref{fig:SQUID}, we present the enhancement examples of deep learning-based methods on images from SQUID dataset. The input underwater images photographed from different water depths (from 3-6 meters to 20-30 meters) present diverse tough problems for all comparison methods. The Water-Net and UICoE-Net remove the haze better, but fail to recover the color. Ucolor slightly improves the color appearance in some cases but fails to deblur and recover the details.  Benefiting from the guidance of high-level semantic information, our approach performs impressive visual enhancement with the extra constraints for these uncommon underwater environments. The experimental results also show that our SGUIE-Net successfully performs color restoration without artifacts and improves the local details and global contrast of the input underwater images. For more visual comparison on the images from SQUID, please refer to the supplement materials.

To quantitatively evaluate the performance of color restoration, we calculate the average reproduction angular error (AE) denoted as $ \overline{\psi}$ between the gray-scale patches and corresponding pure gray colors in RGB space. We use the evaluation code provided by Berman et al. \cite{berman2020underwater} and evaluate the performance of the deep learning-based methods. Figure \ref{fig:SQUID_h} shows the AE scores for each color card in Figure \ref{fig:SQUID}, demonstrating that our results not only provide a pleasing visual perception, but also receive quite good color restoration accuracy.
Full average evaluation results on all the 57 test images are shown in Table \ref{tab:SQUID}. 
Our SGUIE-Net achieves the lowest mean angular error on all test images. In addition, our method improves more than 40\% in terms of PS compared to the second place, which is consistent with qualitative analysis of the visual comparison.

The Color-Checker7 dataset contains 7 underwater images taken from a shallow swimming pool with different cameras. Color checker is also photographed in each image. It provides a good path to demonstrate the robustness of our method to different imaging devices and the accuracy of color restoration. We follow Ancuti et al. \cite{Ancuti2018Color} to employ CIEDE2000\cite{sharma2005ciede2000} to measure the relative differences between the corresponding color patches of ground-truth Macbeth Color Checker and the enhancement results of these comparison methods.  As shown in Figure \ref{fig:ColorCheck7}, the professional underwater camera (Pentax W60 and Olympus T6000) also inevitably introduces various color casts. Although Water-Net removes the bluish distortion of the input images, it still leaves the problems of low contrast and excessive reddish compensation. Our SGUIE-Net, UICoE-Net and Ucolor perform better than Water-Net and do not introduce obvious artificial colors. In addition, we report the CIEDE2000 scores of all comparison methods in each image and gives the average score subsequently in Table \ref{tab:ColorCheck7}. Our SGUIE-Net achieves the best performance on four cameras, and receives very competitive scores on the other three cameras, resulting in the best average score in the end, which also proves that our method also achieves impressive accuracy of color restoration.

\subsection{Ablation Study}

\begin{figure}[t]
  \begin{center}
  \includegraphics[width=1\linewidth]{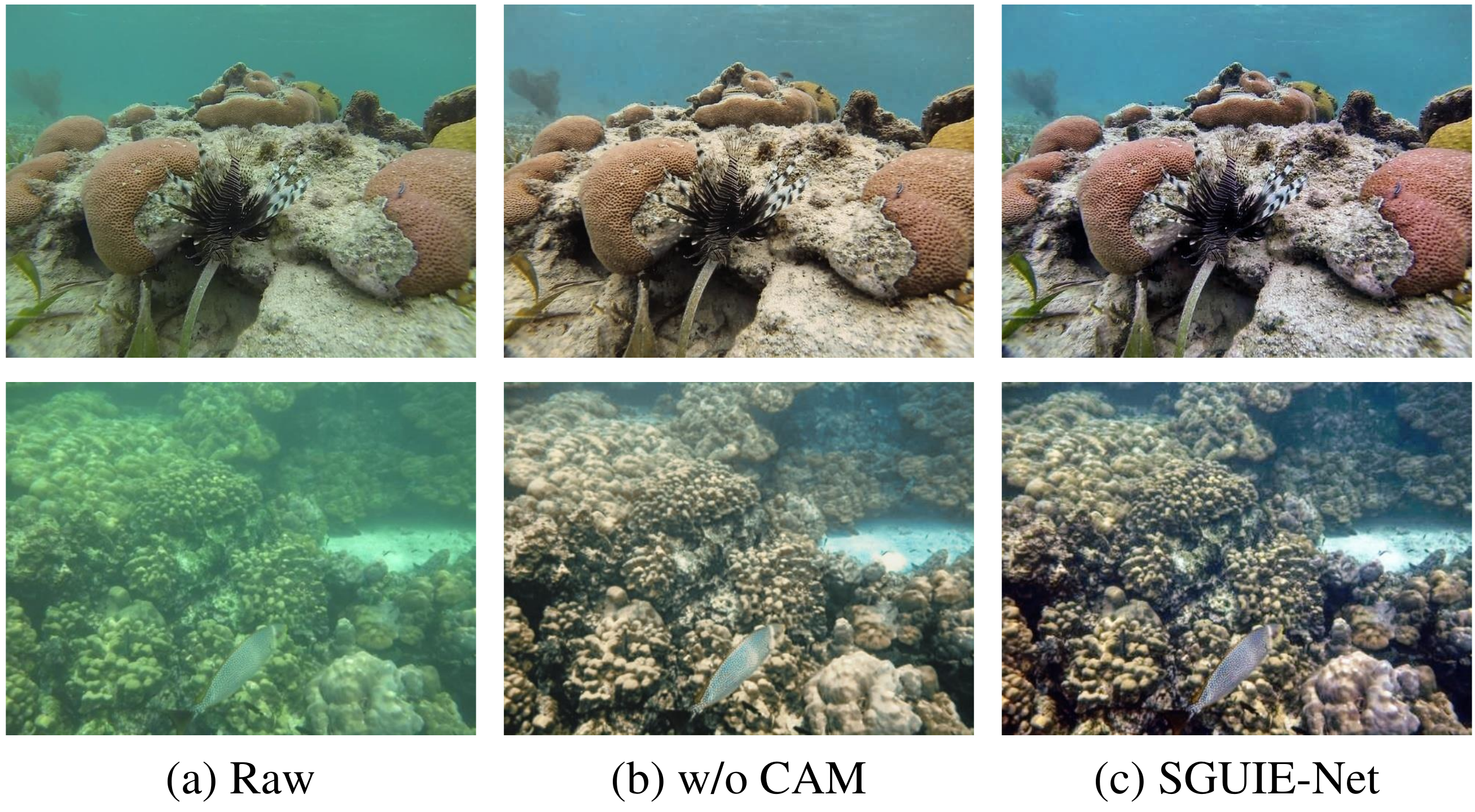}
  \caption{Ablation study of the contributions of the cascaded attention-aware enhancement module (CAM). From left to right are raw images and enhancements with the SGUIE-Net-w/o-CAM and complete SGUIE-Net. With CAM, SGUIE-Net recovers more local details.}
  \label{fig:Ablation_CAM}
  \end{center}
\end{figure}

To demonstrate the effect of the core components in our network, we conduct a series of ablation studies involving the following experiments: 
\begin{itemize}
  \item SGUIE-Net without cascaded attention-aware enhancement module (w/o CAM)
  \item SGUIE-Net without semantic region-wise enhancement module (w/o SRM)
  \item SGUIE-Net using SRM without semantic-based region split (w/o SS)
  \item SGUIE-Net using SRM without U-Net block (w/o UNet)
  \item SGUIE-Net without feature attention block (w/o FAB)
  \item SGUIE-Net using FAB without channel attention (w/o CA)
  \item SGUIE-Net using FAB without pixel attention (w/o PA)
\end{itemize}

The SSIM, PSNR and MSE scores on SUIM-E and UIEB datasets are given in Table \ref{tab:Eva_AB1}. On both two test datasets, our full model achieves the best performance compared to all the ablation models, which also proves the effectiveness of the CAM, SRM and FAB modules.
The CAM module is designed without down-sampling or up-sampling operations to preserve the detailed texture appearance of the input image and thus obtain a better visual quality. Figure \ref{fig:Ablation_CAM} shows the visual comparison between the ablation model w/o CAM and the full SGUIE-Net, demonstrating the important role of the CAM module in complementing detailed information.

\begin{figure}[t]
  \begin{center}
  \includegraphics[width=1\linewidth]{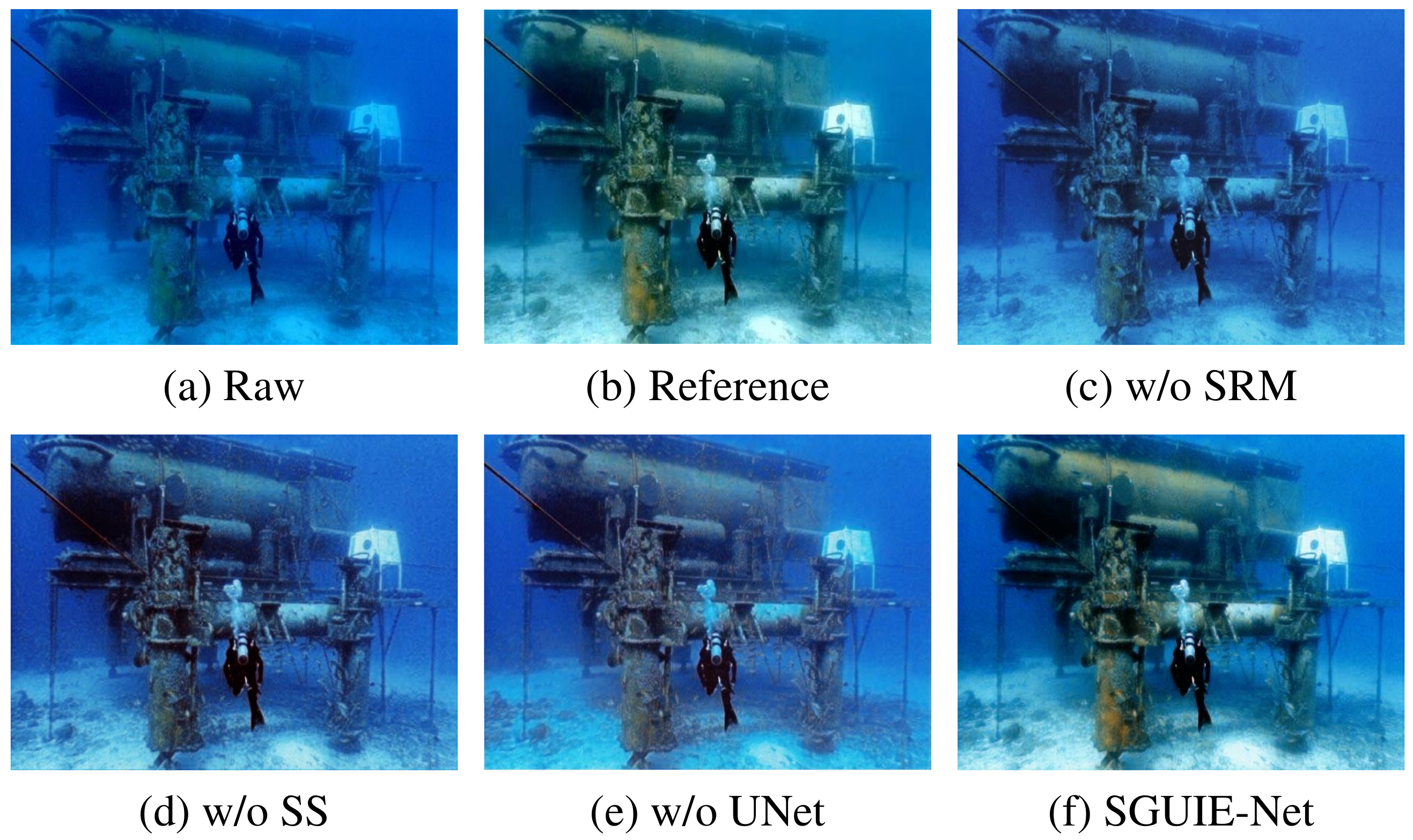}
  \caption{Ablation study of the contributions of the semantic region-wise enhancement module. From (a) to (f) are raw image, enhancement reference and enhancements with SGUIE-Net by removing SRM, SGUIE-Net with random split regions enhancement block, SGUIE-Net by removing U-Net block and the complete SGUIE-Net. SGUIE-Net achieves better color correction.}
  \label{fig:Ablation_Network}
  \end{center}
\end{figure}

The SRM module is a key component of our network, through which our network can capture multi-scale perception with semantic relations across different images to build high-level enhancement guidance. Compared with the SGUIE-Net-w/o-SRM, our complete SGUIE-Net with SRM module improved SSIM and PSNR scores with nearly 4\% and 13\%, respectively, and the MSE is reduced by 39\% on SUIM-E. Specially, for the SRM module, in order to fully verify that the semantic-based region enhancement is meaningful, we also tested the enhancement by randomly dividing the regions. Compared to the complete SGUIE-Net with semantic-based region split strategy, SGUIE-Net-w/o-SS reduces the performance from 0.928, 24.820 and 307 to 0.907, 22.841 and 418 in SSIM, PSNR and MSE metrics, respectively. This also demonstrates that proper semantic guidance can help the enhancement model to generate better results. The visual comparisons shown in Figure \ref{fig:Ablation_Network} can also significantly support the above points. Compared with its ablated versions, the proposed SGUIE-Net provides better color correction and visually pleasing enhancements, with the guidance of high-level semantic information. In contrast, SGUIE-Net-w/o-SRM and SGUIE-Net-w/o-SS are unable to adaptively apply differentiated enhancement to semantic regions with different degrees of degradation due to the lack of semantic region awareness (see Figure \ref{fig:Ablation_Network}(c) and (d)). Without multi-scale perception, SGUIE-Net-w/o-UNet can only introduce local semantic perception, whose enhancement shows as local detail enhancement but not global improvement (see Figure \ref{fig:Ablation_Network}(e)). Besides, the reduced performance of SGUIE-Net-w/o-SRM, SGUIE-Net-w/o-CAM and SGUIE-Net-w/o-UNet further demonstrate the effectiveness and rationality of the dual-branch, multi-scale feature perception architecture.

\begin{figure}[t]
  \begin{center}
  \includegraphics[width=1\linewidth]{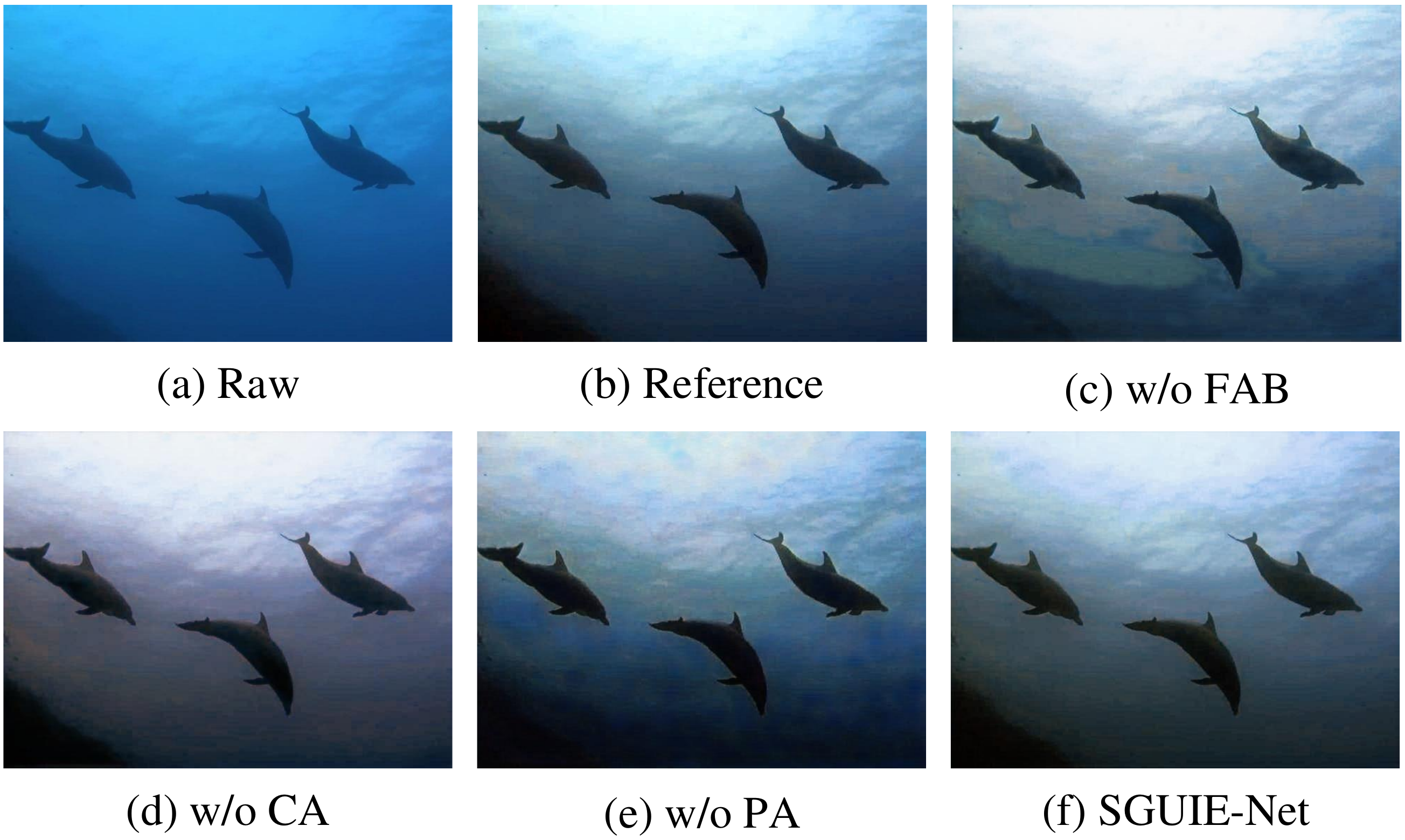}
  \caption{Ablation study of the effectiveness of the channel attention module and pixel attention module. From (a) to (f) are raw image, enhancement reference and enhancements with SGUIE-Net by removing feature attention block, SGUIE-Net by removing channel attention module, SGUIE-Net by removing pixel attention module and the complete SGUIE-Net.}
  \label{fig:Ablation_FAB}
  \end{center}
\end{figure}

The FAB module, as the basic component of our network, contains the cascaded channel attention block and pixel attention block. To demonstrate the effectiveness of these two attention blocks, we also performed ablation experiments. Numerically, the PSNR score of SGUIE-Net-w/o-PA is higher than that of SGUIE-Net-w/o-CA by about 0.18, and both of these two ablation models are better than SGUIE-Net-w/o-FAB. However, their best scores are still lower than our full version of the method in all SSIM, PSNR, and MSE metrics. Figure \ref{fig:Ablation_FAB} shows the different ablation results of the FAB module. Due to the lack of weighted perception of the channel attention block for different feature components, only using the pixel attention module may introduce artificial colors into the enhancement results (see Figure \ref{fig:Ablation_FAB}(d)). While, using the channel attention block alone ignores the spatially uneven degradation of different pixels, resulting in unnatural local color distortion (see Figure \ref{fig:Ablation_FAB}(e)). Moreover, the visual quality and quantitative evaluation of SGUIE-Net-w/o-FAB are disappointing. Both qualitative and quantitative results can demonstrate that the FAB effectively combine the advantages of the two attention blocks to achieve better performance.


     

\section{Conclusion}\label{sec:Conclusion}
In this paper, we propose a novel underwater image enhancement network with semantic attention guidance and multi-scale perception.
By introducing the semantic information as high-level guidance, we design a semantic region-wise enhancement module to bridge the gap between uncommon degradation  types and the learned distribution of underwater degradation. With the attention mechanism, our model could sense the uneven degradation of different semantic regions and feed back to the global attention features through the embedded semantic guided feature extraction and fusion module. Besides, the proposed complementary dual-branch, multi-scale feature perception architecture allows the model to obtain good global enhancement while recovering clear local details. We conducted extensive experiments on four real-world underwater benchmarks to verify the effectiveness and the color restoration accuracy of our network. Ablation studies further demonstrated the effectiveness of the proposed cascaded attention-aware enhancement, semantic region-wise enhancement and the new dual-branch, multi-scale feature perception architecture.
%


%

%


\ifCLASSOPTIONcaptionsoff
  \newpage
\fi



\bibliographystyle{IEEEtran}
\bibliography{IEEEabrv,reference}
%

%






\end{document}